\documentclass[aps,prd,groupedaddress,showpacs]{revtex4}

\usepackage[parfill]{parskip}    
\usepackage{graphicx}
\usepackage{amssymb}
\usepackage{epstopdf}
\usepackage{color}
\usepackage{float}
\usepackage{amsmath}
\usepackage{inputenc}

\usepackage{graphicx}
\usepackage{pdfpages}
\usepackage[colorlinks=true,citecolor=blue,urlcolor=magenta,breaklinks]{hyperref}

\begin{document}

\title{Durgapal IV model in light of the minimal geometric deformation approach}

\author{Francisco Tello-Ortiz}
\email{francisco.tello@ua.cl
 } \affiliation{Departamento de F$\acute{i}$sica, Facultad de ciencias b$\acute{a}$sicas,
Universidad de Antofagasta, Casilla 170, Antofagasta, Chile.}

\author{\'Angel Rinc\'on}
\email{angel.rincon@pucv.cl}
\affiliation{Instituto de F\'isica, Pontificia Universidad Cat\'olica de Valpara\'iso, Avenida Brasil 2950, Casilla 4059, Valpara\'iso, Chile.}

\author{Piyali Bhar}
\email{piyalibhar90@gmail.com}
\affiliation{Department of Mathematics,Government General Degree College, Singur, Hooghly, West Bengal 712 409, India}

\author{Y. Gomez-Leyton}
\email{ygomez@ucn.cl}
\affiliation{Departamento de F\'isica, Universidad Cat\'olica del Norte, Av. Angamos 0610, Antofagasta,
Chile.}

\begin{abstract}
The present article is devoted to the study of local anisotropies effects on the Durgapal's fourth model in the context of gravitational decoupling via the Minimal Geometric Deformation approach. To do it, the most general equation of state relating the components of the $\theta$--sector is imposed to obtain the decoupler function $f(r)$. In addition, certain properties of the obtained solution are investigated, such as the behavior of the salient material content threading the stellar interior, causality and energy conditions, hydrostatic balance through modified Tolman--Oppenheimer--Volkoff conservation equation and stability mechanism against local anisotropies by means of adiabatic index, sound velocity of the pressure waves, convection factor and Harrison--Zeldovich--Novikov procedure, in order to check if the model is physically admissible or not. Regarding the stability analysis, it is found that the model presents unstable regions when the sound speed of the pressure waves and convection factor are used in distinction with what happens in the adiabatic index and Harrison--Zeldovich--Novikov case. To produce a more realistic picture the numerical data for some known compact objects was placed and different values of the parameter $\alpha$ were considered to compare with the GR case \i.e, $\alpha=0$. 
\end{abstract}


\maketitle

\section{Introduction}
The study of compact structures within the framework of Einstein gravity theory \i.e., General Relativity compromises a great challenge. In this regard, the simplest situation is the study of spherically symmetric and static configurations made up of perfect fluid distributions that is, equal radial and tangential pressures $p_{r}=p_{t}=p$.
Such solutions have been extensively investigated, starting from the seminal work of Tolman \cite{r82}, up to current works \cite{Panotopoulos:2019wsy,Panotopoulos:2019zxv,Panotopoulos:2020zqa}, just to mention a few of them.
The problems in consider the above situation are: i) Not all of these solutions meet the requirements in order to represent a well established physical model \cite{delgaty} and ii) from the astrophysical point of view a perfect fluid matter distribution does not represent the most real situation. Relaxing the condition $p_{r}=p_{t}$ and allow the presence of local anisotropies intrigues a more exciting and realistic picture. The anisotropic behavior of matter distributions is characterized by the so called anisotropy factor $\Delta\equiv p_{t}-p_{r}$. This parameter quantifies the deviation of the fluid pressures waves in the principal directions \i.e, the radial and transverse directions of the fluid sphere. The effects introduced by an anisotropic material content on the main features of stellar interiors are well understood \cite{bowers,r1,r2,r3,r4,r5,r6,r7,r8,r9,r10,r11,r12,r13,r14,r15,r16,r17,r18,r19,r20,r21,r22,r23,Lopes:2019psm,Tello-Ortiz:2020svg}. In this concern the plethora of works available in the literature provide at the theoretical level a better understanding of the impact of this imperfect fluid distributions on compact configurations, which allows to contrast with astrophysical observations \cite{r24,r25,n1,r26,r27,r28,r29,r30,r31,n2,n3,n4,n5,r32,r33,r34,r35,n6,n7,n8,n9} (and references contained therein).

Although the study of anisotropic compact structures is abundant, a very important and interesting point is how to produce the anisotropic behavior inside the compact structure. Over the last decade a simple and powerful methodology to extend isotropic solution of Einstein field equations to an anisotropic domains was developed \cite{r36,r37,r38,r39,r40,r41,r42,r43,r44,r45,r46}. This procedure called gravitational decoupling by \emph{minimal geometric deformation} (MGD hereinafter) has gain many adepts these two last years. The subjects covered by this approach involve anisotropic neutron stars, black hole solutions, modified and alternative gravity theories, cosmology, among others \cite{r48,r49,r50,r51,r52,r53,r54,r55,r56,r57,r58,r60,r61,r62,r63,r64,r65,r66,r67,r68,r69,r70,r71,r72,r73,r74,r75,r76,r77,Rincon:2020izv,Abellan:2020jjl}. The main points of this methodology are the extension of the energy-momentum tensor to a more complicated one given by
\begin{equation}
T_{\mu\nu}=\tilde{T}_{\mu\nu}+\alpha\theta_{\mu\nu},    
\end{equation}
where $\tilde{T}_{\mu\nu}$ corresponds to the seed energy-momentum tensor (in the simplest case it represents a perfect fluid distribution), $\theta_{\mu\nu}$ an extra piece which introduces the anisotropic behaviour and $\alpha$ a dimensionless parameter. The second main ingredient is the mapping of one of the metric potentials $e^{\nu}$ or $e^{\lambda}$ (usually the $g^{rr}$ metric component is deformed) to
\begin{equation}
e^{-\lambda(r)} \mapsto \mu(r)+\alpha f(r),    
\end{equation}
being $f(r)$ the so called decoupler or minimal geometry deformation function. These ingredients with the original solution of Einstein field equations constitute the extended or deformed solution. 

So, following the same spirit of \cite{r46}, in this work we extend Durgapal's fourth model \cite{durg} to an anisotropic scenario. To determine the full $\theta$-sector and the minimal geometric deformation function $f(r)$ we implement an equation of state (EoS from now on) relating the components of the extra piece $\theta_{\mu\nu}$. This EoS connects $\theta^{t}_{t}$, $\theta^{r}_{r}$ and $\theta^{\varphi}_{\varphi}$ by means of two free constant parameters, namely $a$ and $b$. Once the $\theta$-sector is fully determined and the decoupler function $f(r)$ is obtained the problem is closed. However, it should be noted that the form of the function $f(r)$ strongly depends on the election of the parameters $a$ and $b$. With the complete geometry and energy-momentum tensor in hand we proceed to study the behaviour of the model at the surface, joining it with the well known Schwarzschild vacuum space-time. Nevertheless, the outer space-time could in principle receive contribution from the $\theta$-sector. So, in that case one is dealing with a non empty space-time. Although this situation could indeed introduce intriguing insights, in this opportunity and without lost of generality it is also valid to use the original Schwarzschild solution. The matching condition scheme allows to us to obtain the mass of the fluid sphere and all the constant parameters that characterize the model. 

In order to mimic a more realistic compact object such as neutron or quark stars, we have fixed the mass and radius corresponding to some known compact objects, specifically Her X--1 \cite{abu}, SMC X--1 and LMC X--4 \cite{rawls}. 
In this direction, recent works available in the literature \cite{sarkar,tello} have addressed the construction of anisotropic fluid spheres determining the maximum possible mass of the compact object by analysing the $M-R$ profile. In \cite{sarkar} was found that the $M-R$ profile matches the numerical data for some known compact objects such as Vela X-1, Cen X-3, EXO 1785-248 and LMC X-4, while in \cite{tello} the $M-R$ curve shows that it is possible to reach masses greater than $3M_{\odot}$. In comparison with these works, in the present one the obtained mass values are closer to the reported values for the aforementioned compact structures.
Regarding the sign and magnitude of the constant parameter $\alpha$, the structure of the model naturally restricts the parameter to be a strictly positive defined quantity \i.e, $\alpha>0$ in order to ensure a positive anisotropy factor $\Delta$, which corresponds to a physical acceptable situation. Besides, to reproduce the salient model we have considered as an example two different values for $\alpha$, namely $-1$ and $-2$, what is more we have discussed and computed the GR case ($\alpha=0$) to contrast with the MGD case. Additionally, we have discussed and analyzed the main salient features of the model as well as the equilibrium and stability mechanism.

The article is organized as follows. In Sec. \ref{section2} we revisited in short the well known Durgapal's model. Sec. \ref{section3} presents gravitational decoupling by means of minimal geometric deformation scheme. In Sec. \ref{section4} the obtained model is presented and a physical analysis is performed. Besides, the junction condition procedure by using Israel-Darmois matching conditions is discussed. Sec. \ref{section5} reports the main results of this study and finally Sec \ref{section6} provides some remarks and conclusions. 

Throughout the study we adopt the mostly negative metric signature $(+, -, -, -)$, and
we choose natural units where $c = 1 = G$, then the gravitational constant $\kappa$ is equal to $8\pi$.

\section{Durgapal's fourth model}\label{section2}
More than 30 years ago M.C. Durgapal noticed that whether we consider a simple relation $\text{e}^{\nu} \propto (1 + x)^n$ (with $n$ a real and positive value), we are able to obtain exact solutions of the Einstein field equations. Also, notice that 
the solutions become physically relevant whether i) the pressure and the density are finite and positive and, ii) the density, $\tilde{p}/\tilde{\rho}$ and $\mathrm{d}\tilde{p}/\mathrm{d}\tilde{\rho}$ decrease as one goes outwards from the centre $(r=0)$ to the surface of the star $(r=R)$ (see \cite{durg} for seminal work).
In this section we will focus on the Durgapal's fourth model, which is described by the following line element \cite{durg}
\begin{align}\label{Durg}
    \mathrm{d}s^2 &=  \text{e}^{\nu}\mathrm{d}t^2 -  \text{e}^{\lambda}\mathrm{d}r^2 -
    r^2\mathrm{d}\Omega^2,
\end{align}
where the functions involved are defined as follow:
\begin{eqnarray}\label{eq2}
 \text{e}^{\nu(x)} &=& A(1 + x)^4,
 \\ \label{eq3}
 \text{e}^{-\lambda(x)} &=& \frac{7-10 x - x^2}{7(1 + x)^2}
 +
 \frac{Bx}{(1 + x)^2 (1 + 5 x)^{2/5}},
 \\ \label{eq4}
 x &=& C r^2,
\end{eqnarray}
Please note that the set of constant $A$ and $B$ are dimensionless parameters whereas the function $x$ is dimensionless too. The parameter $C$ has units of length$^{-2}$. The later space-time describes a spherically symmetric and static configuration associated to an isotropic matter distribution i.e. equal radial and transverse pressures $p_{r}=p_{t}$. The perfect fluid parameters that characterize this model are given by
\begin{align}
    \tilde{\rho}(r) & = \frac{C}{\kappa}
    \Bigg[
    \frac{8}{7}\frac{(9 + 2x + x^2)}{(1+x)^3}-\frac{B 
    (3 + 10x - 9x^2)}{(1+x)^3 (1 + 5x)^{ \frac{7}{5} }}
    \Bigg], \label{isodensity}
    \\
    \tilde{p}(r) &=  \frac{C}{\kappa}
    \Bigg[ \frac{16}{7}\frac{(2 - 7x - x^2)}{(1+x)^3} +
    \frac{B (1 + 9x)}{(1+x)^3 (1 + 5x)^{ \frac{2}{5} } }
    \Bigg], \label{isopressure}
\end{align}
As was pointed out earlier, one of the essential features of anisotropic models is the inequality between radial and tangential pressures i.e. $p_{r}\neq p_{t}$. In the next section we will introduce the mechanism to include anisotropies in the above model. It is worth mentioning that this model was already worked in the context of anisotropic fluid distributions and charged ones \cite{sunil1,sunil2,sunil3}, obtaining, a well posed toy model to describe compact structures. As we will see later, these antecedents shall be used to compared some aspects of the resulting to model provided by this article.

\section{Minimal geometric deformation and gravitational decoupling} \label{section3}
This section is introduce the well-known minimal geometric deformation (MGD hereafter) approach. The later is a novel tool useful to generate anisotropic solutions of the Einstein field equations starting from a isotropic one
\cite{r46}.
In general, there are many ways to introduce anisotropies in isotropic (anisotropic) models, however,  we will only focus on the case where the shear term (of the energy--momentum tensor) is taken to be zero. Thus, the inclusion of (additional) anisotropies considered here appears when $p_t - p_r \neq 0$. %
In order to produce the aforementioned anisotropies, it is necessary to account a supplementary gravitational source which, in principle, could be e.g. a tensorial, a vectorial or an scalar field. 
The latter source is considered as an additional term to the energy--momentum tensor associated to the seed solution. The coupling appears via a dimensionless coupling constant $\alpha$. The effective energy momentum tensor can be defined as follows
\begin{equation}\label{effectivestresstensor}
{T}_{\mu\nu} \equiv \tilde{T}_{\mu\nu} + \alpha \theta_{\mu\nu},
\end{equation}
where $\tilde{T}_{\mu\nu}$ is associated to a perfect fluid given by (\ref{isodensity})-(\ref{isopressure}), and $\theta_{\mu \nu}$ encoded the anisotropies contributions. Now, taking advantage of the expressions (\ref{Durg}) and (\ref{effectivestresstensor}) the effective Einstein field equations are
\begin{eqnarray}
\label{effectivedensity}
\kappa {\rho}&=&\frac{1}{r^2}-e^{-\lambda}\left(\frac{1}{r^2}-\frac{\lambda^{\prime}}{r}\right),
\\ 
\label{effectiveradialpressure}
\kappa {p}_{r}&=&-\frac{1}{r^2}+e^{-\lambda}\left(\frac{1}{r^2}+\frac{\nu^{\prime}}{r}\right),
\\
\label{effectivetangentialpressure}
\kappa {p}_{t}&=&\frac{1}{4}e^{-\lambda}\left(2\nu^{\prime\prime}+\nu^{\prime2}-\lambda^{\prime}\nu^{\prime}+2\frac{\nu^{\prime}-\lambda^{\prime}}{r}\right),
\end{eqnarray}
where primes mean differentiation with respect to the radial coordinate $r$.  The Bianchi's identities are given by 
\begin{equation}\label{bianchi}
\nabla_{\mu}T^{\mu}_{\nu}=0,
\end{equation}
which reads
\begin{align}
\begin{split} \label{energyfull}
   & \tilde{p}^{\prime} + \frac{1}{2}\nu^{\prime} \left(\tilde{p} + \tilde{\rho}\right) -
   \alpha  K(\theta_i^i)
=0,
\end{split}
\end{align}
where the function $K(\theta_i^i)$ encodes the corresponding anisotropies and it is defined as
\begin{align} \label{Ktheta}
    K(\theta_i^i) \equiv  \left(\theta^{r}_{r}
    \right)^{\prime}  + \frac{1}{2}\nu^{\prime} \left(\theta^{r}_{r}-\theta^{t}_{t} \right) + \frac{2}{r} \left(\theta^{r}_{r}-\theta^{\varphi}_{\varphi}\right),
\end{align}
being the Eq.~\eqref{energyfull} a linear combination of the effective pressures and density. Thus, the parameters involved are: i) the effective density ${\rho}$, ii) the effective radial pressure ${p}_{r}$ and finally iii) the effective tangential pressure ${p}_{t}$. 
We link the above functions with the perfect fluid parameters as follows
\begin{eqnarray}\label{effecrho}
{\rho}&\equiv&\tilde{\rho}+\alpha \theta^{ t}_{t}\\\label{effecpr}
{p}_{r}&\equiv&\tilde{p}-\alpha \theta^{r}_{r}\\ \label{effecpt}
{p}_{t}&\equiv& \tilde{p}-\alpha \theta^{ \varphi}_{\varphi}.
\end{eqnarray}
It is essential to point out that the inclusion of the $\theta$-term introduces anisotropies if $\theta^{r}_{r}\neq \theta^{\varphi}_{\varphi}$ only. Thus the effective anisotropy is defined in the usual manner, namely
\begin{equation}\label{anisotropy}
\Delta\equiv {p}_{t}-{p}_{r}=\alpha\left(\theta^{r}_{r}- \theta^{\varphi}_{\varphi}\right).
\end{equation}
Naturally, we recover a perfect fluid when $\alpha$ is taken to be zero.
In general, it is not trivial to obtain analytic solutions of the Einstein field equations in the context of interior solutions \i.e, relativistic stars. To find a tractable exact solution, albeit recent, a popular alternative is the gravitational decoupling via the MGD approach.
The crucial point of this technique relies on the following maps
\begin{eqnarray}\label{deformationnu}
\text{e}^{\nu(r)}&\mapsto& e^{\nu(r)}+\alpha h(r) \\ \label{deformationlambda}
\text{e}^{-\lambda(r)}&\mapsto& \mu(r)+\alpha f(r),
\end{eqnarray}
in which we deform minimally the $g_{tt}$ and $g_{rr}$ components of the metric. The later maps deform the metric components by the inclusion of certain unknown functions $h(r)$ and $f(r)$.
At this level, it is noticeable remark that the corresponding deformations are purely radial. The later feature remains the spherical symmetry of the solution. The Minimal Geometric Deformation method corresponds to set $h(r)=0$ (with $f(r)\neq 0$, or $h(r)\neq 0$ with $f(r)=0$). The first case maintain the deformation in the radial component only, which means that any temporal deformation is excluded. In light of this, the anisotropic tensor $\theta_{\mu\nu}$ is produced by the radial deformation (\ref{deformationlambda}).

The system of differential equations can be split under the replacement (\ref{deformationlambda}). Thus, the field equations are naturally decoupled in two set: i) the first set satisfies Einstein field equations, and correspond to the isotropic (anisotropic) case, namely $\alpha = 0$, and it is given by
\begin{eqnarray}\label{ro1}
\kappa\tilde{\rho}&=&\frac{1}{r^{2}}-\frac{\mu}{r^{2}}-\frac{\mu^{\prime}}{r}\\\label{p1}
\kappa \tilde{p}&=&-\frac{1}{r^{2}}+\mu\left(\frac{1}{r^{2}}+\frac{\nu^{\prime}}{r}\right)\\\label{p2}
\kappa \tilde{p}&=&\frac{\mu}{4}\left(2\nu^{\prime\prime}+\nu^{\prime2}+2\frac{\nu^{\prime}}{r}\right)+\frac{\mu^{\prime}}{4}\left(\nu^{\prime}+\frac{2}{r}\right),
\end{eqnarray}
along with the following conservation equation
\begin{equation}
 \tilde{p}^{\prime} + \frac{1}{2}\nu^{\prime} \left(\tilde{p} + \tilde{\rho}\right)=0,
\end{equation}
and ii) the second set of equations corresponds to the $\theta$-sector which is obtained when we turn on $\alpha$. Thus, the equations for the later sector are:
\begin{eqnarray}\label{cero}
\kappa \theta^{t}_{t}&=&-\frac{f}{r^{2}}-\frac{f^{\prime}}{r}
\\ \label{one}
\kappa \theta^{r}_{r}&=&-f\left(\frac{1}{r^{2}}+\frac{\nu^{\prime}}{r}\right)  \\  \label{dos}
\kappa \theta^{\varphi}_{\varphi}&=&-\frac{f}{4}\left(2\nu^{\prime\prime}+\nu^{\prime2}+2\frac{\nu^{\prime}}{r}\right)-\frac{f^{\prime}}{4}\left(\nu^{\prime}+\frac{2}{r}\right).
\end{eqnarray}
The corresponding conservation equation associated to the $\theta$-sector is computed to be
\begin{equation}\label{conservationtheta}
\left(\theta^{r}_{r}\right)^{\prime}-\frac{1}{2}\nu^{\prime}\left(\theta^{t}_{t}-\theta^{r}_{r}\right)-\frac{2}{r}\left(\theta^{\varphi}_{\varphi}-\theta^{r}_{r}\right)=0.
\end{equation}
It is important to remark that the above equation is precisely the essential point to use the MGD approach, given that it guarantees that the sources $\tilde{T}_{\mu\nu}$ and $\theta_{\mu\nu}$ interact only gravitationally.

\section{The Model}\label{section4}
To solve the system of equations (\ref{cero})-(\ref{dos}) we have considered the following relation between the components of the $\theta$ source,
\begin{equation}\label{eos}
\theta^{t}_{t}=a\theta^{r}_{r}+b\theta^{\varphi}_{\varphi},
\end{equation}
where $a$ and $b$ are dimensionless parameters. 
The above constraint on the components of the anisotropic sector has been recently applied to some problems in the context of black hole physics \cite{r55}. It is worth mentioning that \eqref{eos} can be thought as an equation of state $F(p,\rho) \equiv 0$ given that it combine the anisotropic ``pressure" with the corresponding anisotropic ``density". In such sense, although it is not clear how the component of $\theta_{\mu \nu}$ are related, it is always natural to impose a general relation between them and then try to take an adequate combination of parameters such as \eqref{eos} serves to obtain well-defined physical solutions. Besides, given that our starting point is an isotropic solution, it is also natural that we try to put some constraints on the anisotropic sector, reason why the equation of state is only related to $\theta_{\mu \nu}$. Finally, the MGD approach also needs a supplementary condition to complete the set of equations. Given that $\theta$-sector is in general unknown make sense to explore different possible combinations on this anisotropic tensor. 

The above equation of state (\ref{eos}) allows us to determine the decoupler function $f(r)$. Specifically (\ref{eos}) leads to the following first order differential equation to $f(r)$
\begin{equation}\label{eq29}
\begin{split} f\left[\frac{1}{r^{2}}-a\left(\frac{1}{r^{2}}+\frac{\nu^{\prime}}{r}\right)-\frac{b}{4}\left(2\nu^{\prime\prime}+\nu^{\prime 2}+\frac{2}{r}\nu^{\prime}\right)\right]  + f^{\prime}\left[\frac{1}{r}-\frac{b}{4}\left(\nu^{\prime}+\frac{2}{r}\right)\right]=0,
\end{split}
\end{equation}
where $\nu$ is given by Eq. (\ref{eq2}). The general solution of Eq. (\ref{eq29}) is
\begin{equation}\label{eq30}
\begin{split}
f(r)=\frac{\left(-5Cbr^{2}+2Cr^{2}-b+2\right)^{\frac{\left(-4ab-6b^{2}+16a+4b+8\right)}{\left(b-2\right)\left(5b-2\right)}}r^{-\frac{2\left(a-1\right)}{b-2}}}{\left(Cr^{2}+1\right)^{2}} D,
\end{split}
\end{equation}
being $D$ an integration constant. In order to avoid a singular behavior when $r\rightarrow 0$, one needs to impose some restrictions on the parameters $a$ and $b$, those are: i) $b\neq 2$, ii) $b\neq 5/2$ and iii) $-2\left(a-1\right)/\left(b-2\right)>0$. In this opportunity we have taken $a$ and $b$ to be free parameters. Concretely, we have chosen $a=-1$ and $b=4$, these values produce the following decoupler function $f(r)$
\begin{equation}\label{eq31}
f(r)=\frac{Dr^{2}}{4\left(9Cr^{2}+1\right)^{2}\left(Cr^{2}+1\right)^{2}}.
\end{equation}
So, the constant $D$ has units of length$^{-2}$. This is so because the decoupler function $f(r)$ must be dimensionless. Moreover, for the sake of simplicity we take $D=C$, then Eq. (\ref{eq31}) becomes
\begin{equation}\label{eq32}
f(r)=\frac{Cr^{2}}{4\left(9Cr^{2}+1\right)^{2}\left(Cr^{2}+1\right)^{2}}.
\end{equation}
It is worth mentioning that different elections on parameters $a$ and $b$ conduce to different expressions for $f(r)$, hence it is not always possible to take $D=C$.

\begin{figure}[H]
\centering
\includegraphics[width=0.38\textwidth]{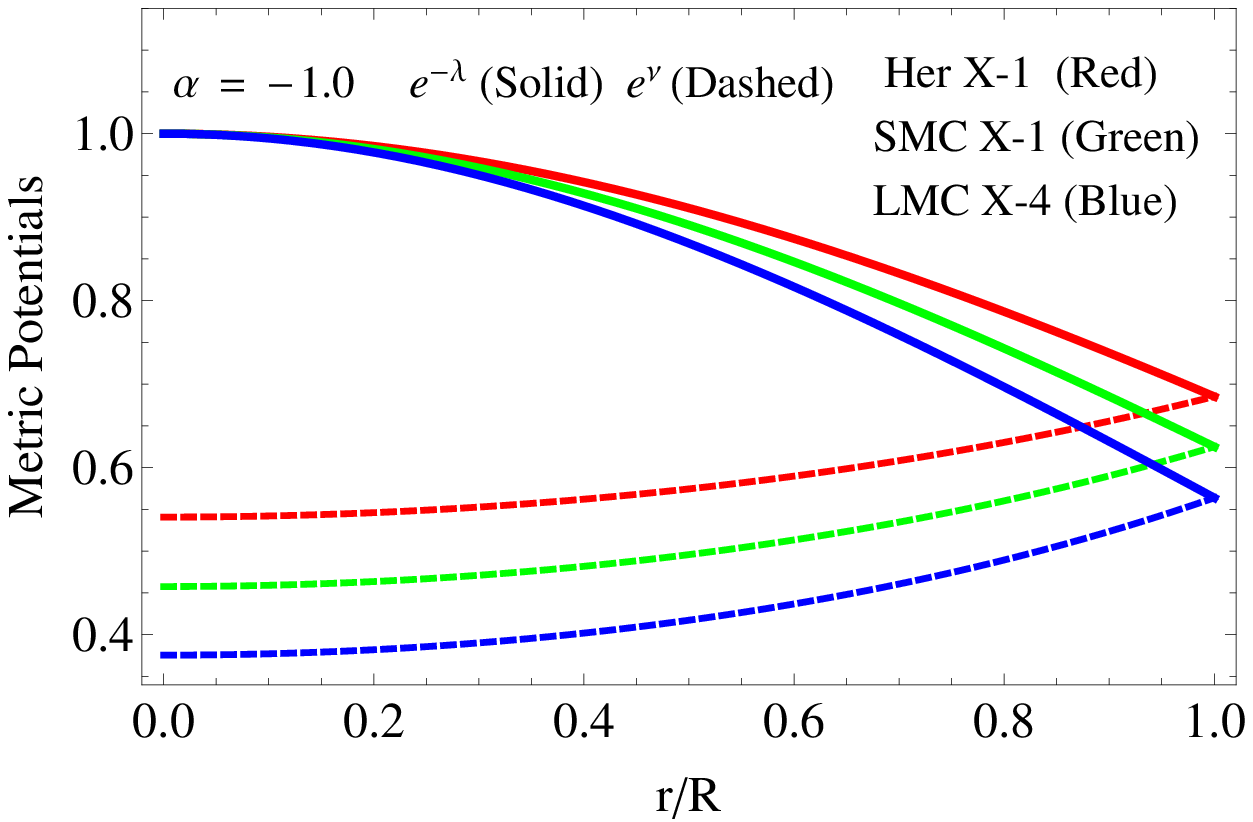} \
\includegraphics[width=0.38\textwidth]{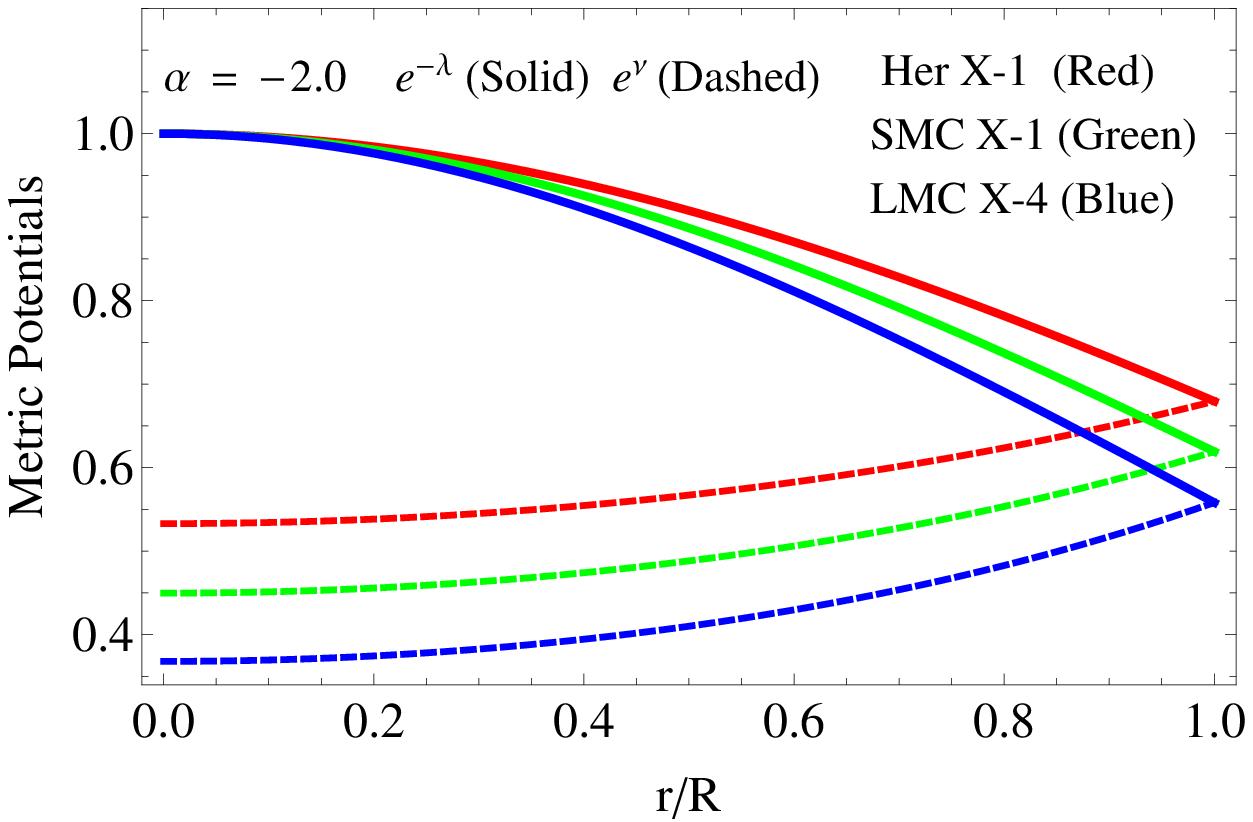}
\caption{Variation of metric potentials with respect to the dimensionless radial coordinate $r/R$, for different values of the constant parameters mentioned in tables \ref{table2} and \ref{table5}.}
\label{fig1}
\end{figure}

Next, by inserting (\ref{eq32}) into (\ref{cero})-(\ref{dos}) we get the following components of the extra field $\theta_{\mu\nu}$ as follows
\begin{eqnarray}\label{eq33}
\theta^{t}_{t}&=&\frac{1}{32\pi}\frac{\left(45C^{2}r^{4}+10Cr^{2}-3\right)C}{\left(9Cr^{2}+1\right)^{3}\left(Cr^{2}+1\right)^{3}}, \\ \label{eq34}
\theta^{r}_{r}&=&-\frac{1}{32\pi}\frac{C}{\left(Cr^{2}+1\right)^{3}\left(9Cr^{2}+1\right)}, \\ \label{eq35}
\theta^{\varphi}_{\varphi}&=&-\frac{1}{32\pi}\frac{\left(9C^{2}r^{4}+2Cr^{2}+1\right)C}{\left(9Cr^{2}+1\right)^{3}\left(Cr^{2}+1\right)^{3}}.
\end{eqnarray}
Thus, the effective thermodynamic variables that quantified the system under study given by (\ref{effecrho})-(\ref{effecpt}) have the following form
\begin{equation}\label{eq36}
\begin{split}
    {\rho}(r)=\frac{C}{8\pi}
    \Bigg[
   \frac{8}{7}\frac{(9 + 2Cr^{2} + C^2r^{4})}{(1+Cr^{2})^3}-\frac{B 
    (3 + 10Cr^{2} - 9C^2r^{4})}{(1+Cr^{2})^3 (1 + 5Cr^{2})^{ \frac{7}{5} }}
    \Bigg]  +\frac{\alpha}{32\pi}\frac{\left(45C^{2}r^{4}+10Cr^{2}-3\right)C}{\left(9Cr^{2}+1\right)^{3}\left(Cr^{2}+1\right)^{3}},
\end{split}
\end{equation}
    \begin{equation}\label{eq37}
    \begin{split}
    {p}_{r}(r)= \frac{C}{8\pi}
    \Bigg[ \frac{16}{7}\frac{(2 - 7Cr^{2} - C^2r^{4})}{(1+Cr^{2})^3} +
    \frac{B (1 + 9Cr^{2})}{(1+Cr^{2})^3 (1 + 5Cr^{2})^{ \frac{2}{5} } }
    \Bigg] 
    +\frac{\alpha}{32\pi}\frac{C}{\left(Cr^{2}+1\right)^{3}\left(9Cr^{2}+1\right)},
\end{split}
\end{equation}
 \begin{equation}\label{eq38}
    \begin{split}
    {p}_{t}(r)= \frac{C}{8\pi}
    \Bigg[ \frac{16}{7}\frac{(2 - 7Cr^{2} - C^2r^{4})}{(1+Cr^{2})^3} +
    \frac{B (1 + 9Cr^{2})}{(1+Cr^{2})^3 (1 + 5Cr^{2})^{ \frac{2}{5} } }
    \Bigg] 
    +\frac{\alpha}{32\pi}\frac{\left(9C^{2}r^{4}+2Cr^{2}+1\right)C}{\left(9Cr^{2}+1\right)^{3}\left(Cr^{2}+1\right)^{3}}.
\end{split}
\end{equation}
Besides from Eq. (\ref{anisotropy}) we get
\begin{equation}\label{eq39}
\Delta(r)=-\frac{\alpha\left(9Cr^{2}+2\right)C^{2}r^{2}}{4\pi\left(9Cr^{2}+1\right)^{3}\left(Cr^{2}+1\right)^{3}}.
\end{equation}
It is worth mentioning that the anisotropy factor (\ref{eq39}) restricts the signature of the $\alpha$ parameter. As can be seen from expression (\ref{eq39}) the whole right member is proportional to $\alpha$ and negative defined unless $\alpha<0$. Therefore, to assure a well defined compact structure, positive $\alpha$ values are discarded in order to avoid nonphysical behaviours into the stellar matter distribution such as instabilities. In what follows we shall consider as an example the values $-1$ and $-2$ and the limit case 0 the contrast the results with the GR scenario. Once the thermodynamic variables are given, the explicit inner geometry obtained with the above formalism describing the deformed Durgapal's IV model is represented by 
\begin{equation}\label{eq40}
\begin{split}
ds^{2}=A\left(1+Cr^{2}\right)^{4}dt^{2}
-\bigg[\frac{7-10Cr^{2}-C^{2}r^{4}}{7\left(1+Cr^{2}\right)^{2}}
+\frac{BCr^{2}}{\left(1+Cr^{2}\right)^{2}\left(1+5Cr^{2}\right)^{2/5}}+\frac{\alpha Cr^{2}}{4\left(9Cr^{2}+1\right)^{2}\left(Cr^{2}+1\right)^{2}}\bigg]^{-1}dr^{2}
-r^{2}d\Omega^{2}.
\end{split}
\end{equation}
At this stage we have the full thermodynamic and geometric description of our model. Next, we explore all the main physical properties that this toy model must fulfill in order to represent compact configurations such as neutron stars, at least form the theoretical point of view.

\subsection{Physical analysis}\label{sub1}
\begin{figure}[H]
\centering
\includegraphics[width=0.32\textwidth]{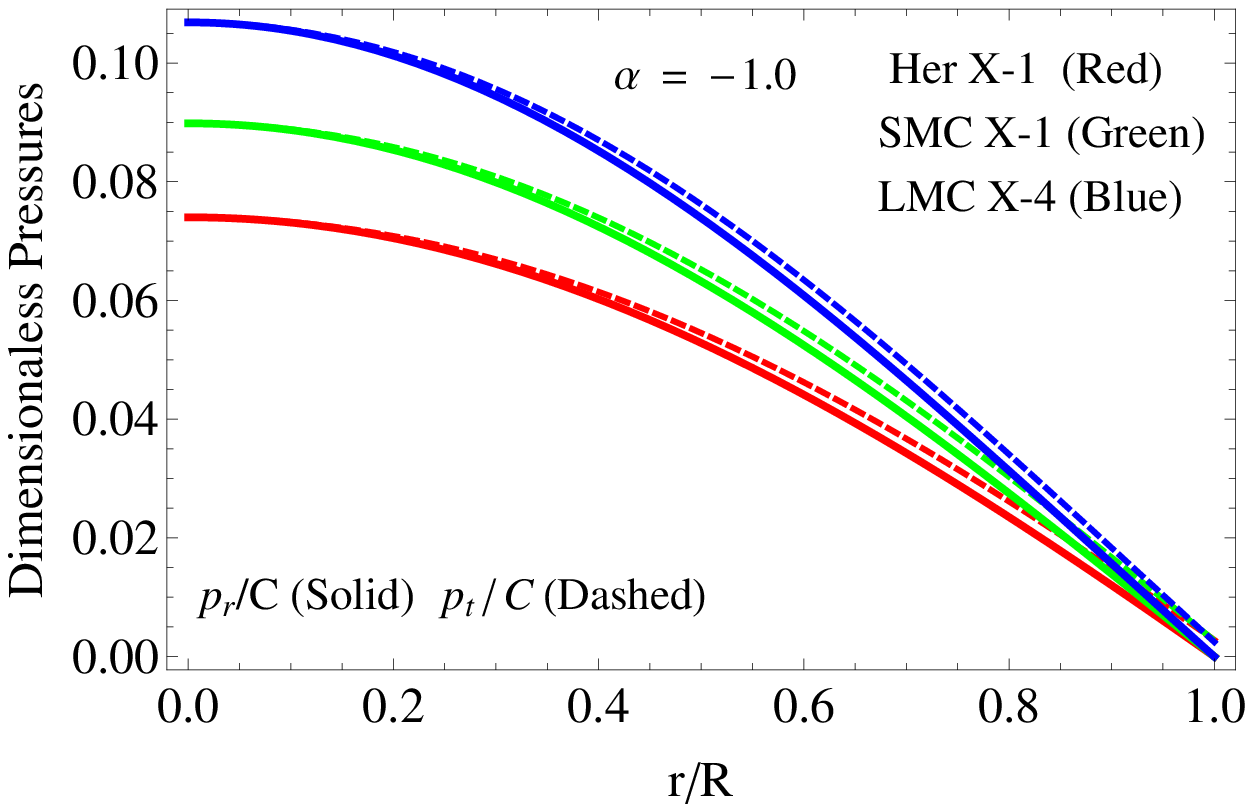}     
\includegraphics[width=0.32\textwidth]{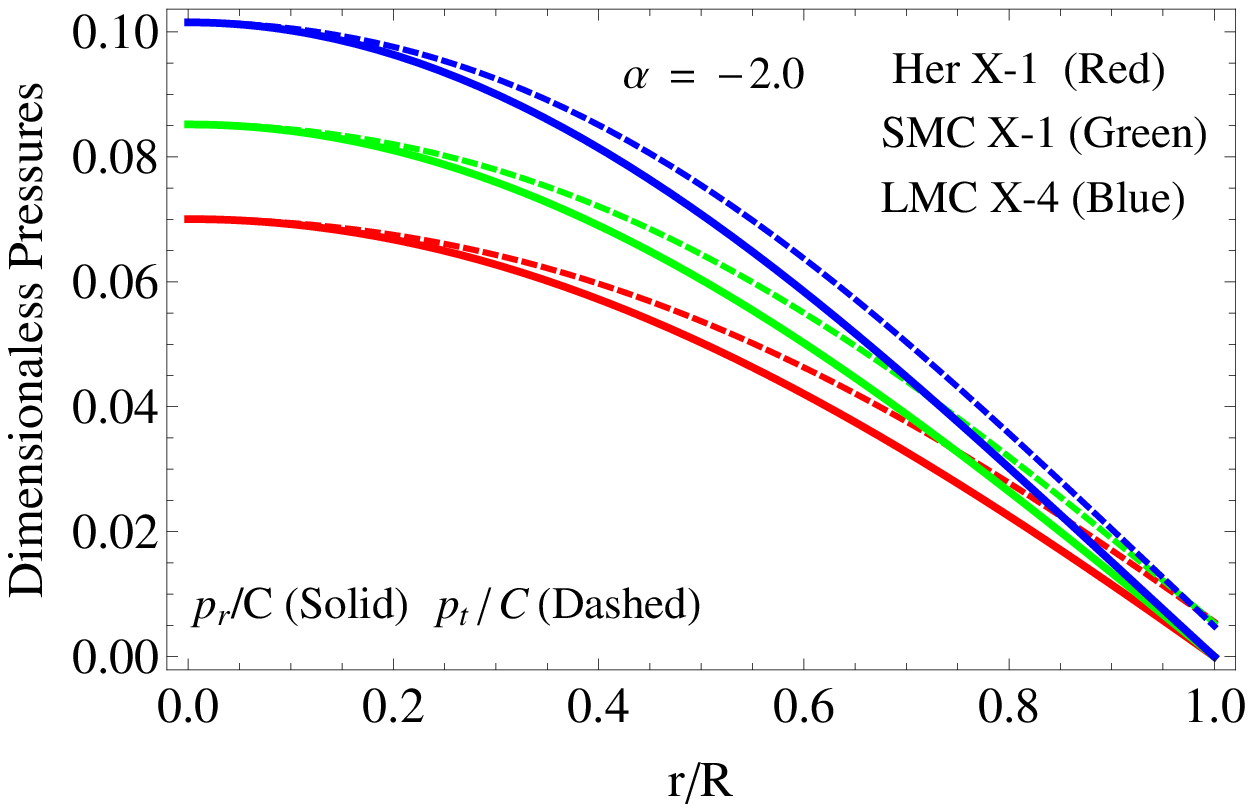}      \\
\includegraphics[width=0.32\textwidth]{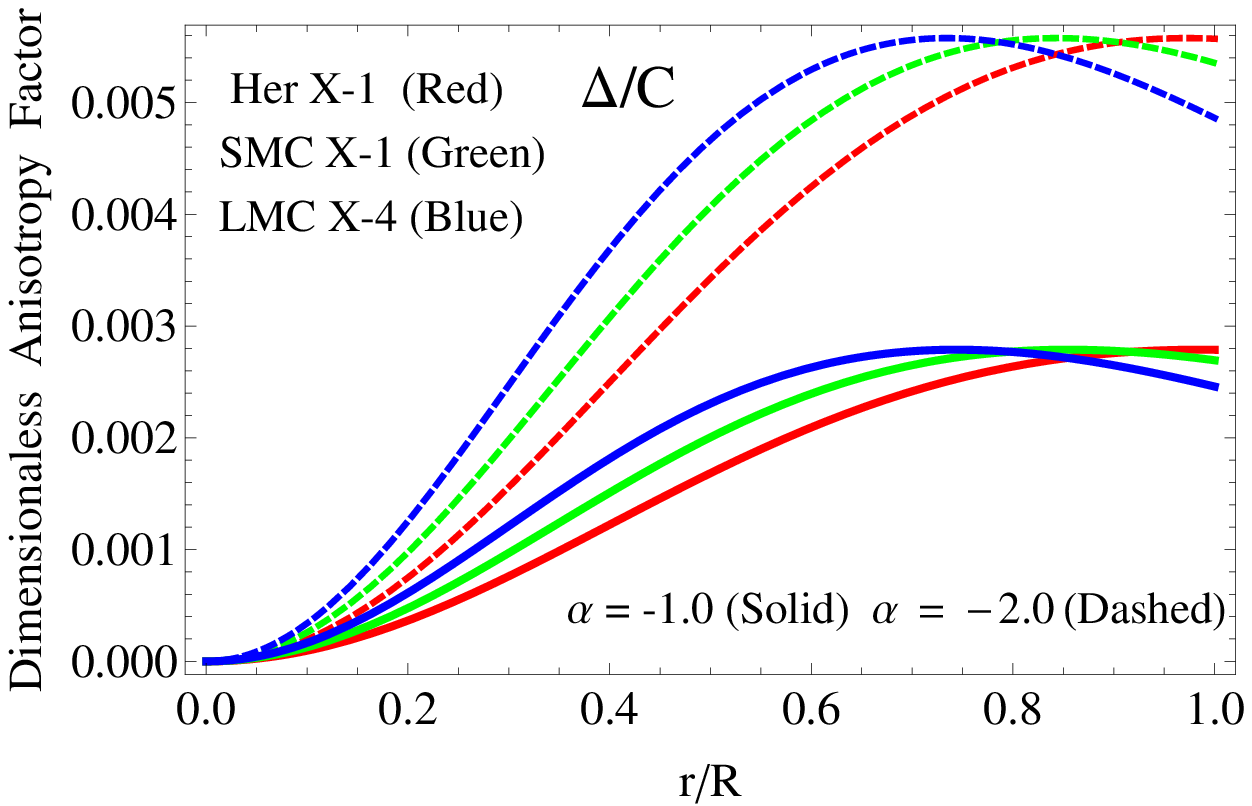}
\includegraphics[width=0.32\textwidth]{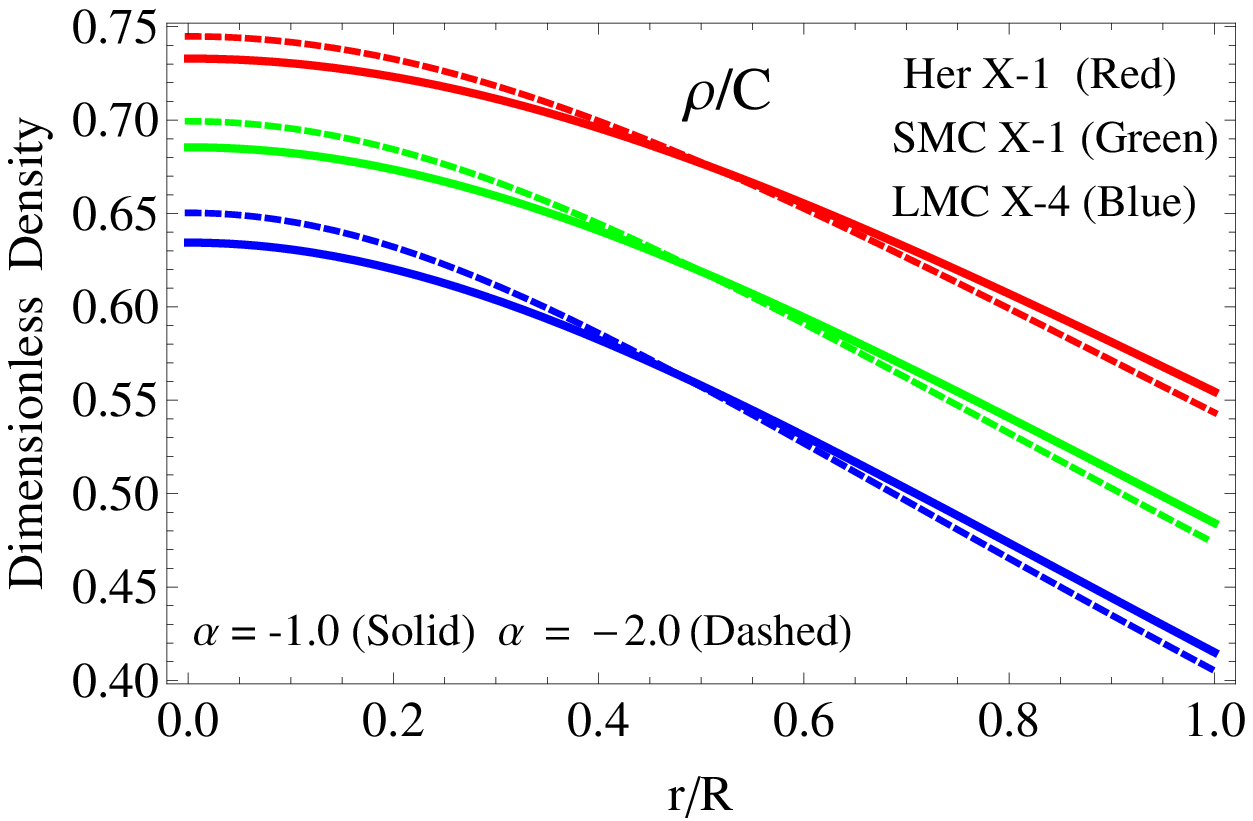}
\caption{
{\bf Upper panels}: The radial $p_{r}$ and tangential $p_{t}$ pressures  against $r/R$. {\bf Lower left panel}: the anisotropy factor $\Delta$ against the dimensionless radius. {\bf Lower right panel} : The energy-density versus the ratio $r/R$. These plots were building by using different values of constant parameters depicted in tables \ref{table2} and \ref{table5}.
}
\label{fig2}
\end{figure}

A well behaved compact structure, describing real celestial bodies, must satisfy certain general requirements. The latter is mandatory to get an appropriate model. 
To check these formalities we explore here in some details the conduct of the main salient features obtained in the previous section, namely the effective energy-density, the tangential and radial pressures, the anisotropy factor and the geometry of the model for all $r$ raging, starting from the center $r=0$, up to the boundary $r=R$. So, regarding the interior manifold it is observed from Eq. (\ref{eq40}) that the pair of metric functions $\{e^{\nu},e^{\lambda}\}$ is completely regular for $r\in[0,R]$. Moreover, $e^{\nu}|_{r=0}>0$, $\left(e^{\nu}\right)^{\prime}|_{r=0}=0$ and $e^{\lambda}|_{r=0}=1$. As we will see later this behavior allows to match the inner geometry to the exterior space-time in a smoothly way at the interface $\Sigma\equiv r=R$ to get the constant parameters that characterize the model. On the other hand, not only the geometry must be well behaved for the interval $0\leq r \leq R$. In this regard the main thermodynamic variables must respect some criteria. From expressions (\ref{eq36})-(\ref{eq38}) at the center of the compact configuration we have
\begin{equation}\label{eq41}
\rho(0)=\frac{C}{8\pi}\left[\frac{72}{7}-3B-\frac{3\alpha}{4}\right],    
\end{equation}
\begin{equation}\label{eq42}
p_{r}(0)=p_{t}(0)=\frac{C}{8\pi}\left[\frac{32}{7}+B+\frac{\alpha}{4}\right].   
\end{equation}
As these quantities should be monotone decreasing functions from the center towards the surface of the star, this entails that their maximum values are attained at $r=0$. What is more these observables must be positive everywhere inside the structure. Thus from (\ref{eq42}) one obtains
\begin{equation}\label{eq43}
-\frac{32}{7}-\frac{\alpha}{4}<B.   
\end{equation}
Furthermore the equation of state parameter $\omega\equiv p_{r}/\rho\leq1$ leads to
\begin{equation}\label{eq44}
B\leq\frac{10}{7}-\frac{\alpha}{4}.
\end{equation}
Therefore the constant parameter $B$ is restricted to lies between
\begin{equation}\label{eq45}
 -\frac{32}{7}-\frac{\alpha}{4}<B\leq\frac{10}{7}-\frac{\alpha}{4}.   
\end{equation}
The monotone decreasing behaviour of these physical quantities means that all of them reach their minimum values at the surface $\Sigma$. In Fig. \ref{fig2} are depicted the trend of the effective radial and tangential pressures. It is observed that the both pressure waves in the principal direction of the object coincide at the center. Besides, the pressure waves in the tangential direction $p_{t}$ dominates the radial one $p_{r}$ at every point inside the star. Furthermore, the effective radial pressure is totally vanishing at the boundary $\Sigma$.  The trace of the effective energy-density $\rho$ is illustrated in the lower row (right panel) of Fig. \ref{fig2}. As can be appreciated its behavior respects the guidelines discussed above for stellar interiors.

\subsection{Junction conditions}\label{sub2}
As compact objects are delimited for the boundary $\Sigma\equiv r=R$, the material content threading the stellar interior must be confined between $0$ and $R$. To ensure the confinement of this matter distribution one needs to study the so called junction conditions at the surface of the structure. In general relativity theory this procedure is carried out by applying the well known Israel-Darmois \cite{r78,r79} matching conditions. To do so the inner manifold $\mathcal{M^{-}}$ describing the stellar interior is matched in a smoothly way at the interface $\Sigma$ with the corresponding exterior manifold $\mathcal{M^{+}}$. In this opportunity the outer space-time is described by the vacuum Schwarzschild solution \cite{r80}
\begin{equation}\label{eq46}
ds^{2}=\left(1-\frac{2M}{r}\right)dt^{2}-\left(1-\frac{2M}{r}\right)^{-1}dr^{2}-r^{2}d\Omega^{2}.
\end{equation}
The above space-time (\ref{eq46}) is pertinent because the present model is representing an uncharged anisotropic solution. Nevertheless, this could involve a more complicated situation due to the modifications introduced in the geometry and material parts by the $\theta$-sector. In principle the new source $\theta_{\mu\nu}$ could modified the exterior geometry as well as its material content. If this is the case the outer manifold (\ref{eq46}) becomes   
\begin{equation}\label{eq47}
ds^{2}=\left(1-\frac{2M}{r}\right)dt^{2}-\left(1-\frac{2M}{r}+\alpha g(r)\right)^{-1}dr^{2}-r^{2}d\Omega^{2},
\end{equation}
where $g(r)$ represents the deformation function (see \cite{r55} and references therein for further details).
In this regard the study of Schwarzschild space-time  minimally deformed has been analyzed in \cite{r54}. In this opportunity and without lost of generality we shall assume that $g(r)=0$, \i.e, the exterior manifold surrounding the stellar interior is the original vacuum Schwarzschild solution. So, the compliance of the matching conditions at the interface $\Sigma$ dictates the continuity of the first and second fundamental forms. At the hyper-surface described by $\Sigma$ the inner manifold $\mathcal{M}^{-}$ and the outer one $\mathcal{M}^{+}$ induce the intrinsic and extrinsic curvatures characterized by the metric tensor $g_{\mu\nu}$ and the symmetric extrinsic curvature tensor $K_{ij}$ (Latin indices run over spatial coordinates only), respectively. 
Given that the metric tensor $g_{\mu\nu}$ is continuous across the surface $\Sigma$, we have, from Eqs. (\ref{eq40}) and (\ref{eq46}) that
\begin{align}\label{eq49}
\begin{split}
 g^{-}_{tt}\Bigl|_{r=R} &= g^{+}_{tt}  \Bigl|_{r=R},
 \\
A\left(1+CR^{2}\right)^{4} &= 1-2\frac{{M}}{R},
\end{split}
\end{align}
\begin{align}\label{eq50}
\begin{split}
&g^{-}_{rr} \Bigl|_{r=R} = g^{+}_{rr} \Bigl|_{r=R} ,
\\
\therefore
&\frac{7-10CR^{2}-C^{2}R^{4}}{7\left(1+CR^{2}\right)^{2}}
+ \frac{BCR^{2}}{\left(1+CR^{2}\right)^{2}\left(1+5CR^{2}\right)^{\frac{2}{5}}}
\\
&+ \frac{\alpha CR^{2}}{4\left(9CR^{2}+1\right)^{2}\left(CR^{2}+1\right)^{2}}
= 1-2\frac{{M}}{R}.
\end{split}
\end{align}
In the expressions (\ref{eq49})-(\ref{eq50}) the quantity ${M}$ which corresponds to the total mass contained by the fluid sphere is replacing the Schwarzschild mass $\tilde{M}$, due to at the boundary $\Sigma$ both quantities coincide \i.e, $m(r)|_{r=R}=M=\tilde{M}$. 
On the other hand, the object $K_{ij}$, which determines the continuity of the components of the second fundamental form, is guaranteed just in case where the surface is free from thin shells or surface layers \cite{r78}.
%
%
In that case the $r-r$ component of $K_{ij}$ leads to 
\begin{equation}\label{eq51}
p_{r}(r)|_{r=R}=0.
\end{equation}
So, the condition $p_r(R)=0$ entails that there is not thin shells. Consequently, the surface density $\sigma$ and pressure $\mathcal{P}$ \cite{r78} induced on the boundary are completely vanished, thus the extrinsic curvature tensor is continuous. To guarantee a null radial pressure at the boundary. From Eq. (\ref{eq51}) one obtains the following relation to determine the constant $B$,
\begin{equation}\label{eq53}
\begin{split}
B=\frac{\left(576R^{6}C^{3}+4096R^{4}C^{2}-704R^{2}C-7\alpha-128\right)}{128\left(81R^{4}C^{2}+18R^{2}C+1\right)}  \left(5R^{2}C+1\right)^{2/5}.
\end{split}
\end{equation}
The continuity of $K_{\theta\theta}$ and $K_{\phi\phi}$ yields to 
\begin{equation}\label{eq544}
M=m(R)=4\pi\int^{R}_{0}\left(\tilde{\rho}+\alpha\theta^{t}_{t}\right) r^{2}dr,  
\end{equation}
being $M$ the total mass contained by the fluid sphere. It should be noted that the expression (\ref{eq50}) is completely equivalent to the expression (\ref{eq544}) \footnote{For more details see appendix \ref{A}.}.
Then the set of Eqs. (\ref{eq49}), (\ref{eq50}) and (\ref{eq53}) are the necessary conditions to obtain all the constant parameters. In following we shall fix $\alpha$, $C$ and the radius $R$. In table \ref{table1} are displayed the obtained mass and the parameters $A$ and $B$. 

\begin{table}[H]
\caption{The numerical values for the total mass, compactness factor and surface gravitational red--shift for $a=-1$, $b=4$ and $\alpha=-1.0$. }
\label{table1}
\begin{tabular*}{\textwidth}{@{\extracolsep{\fill}}lrrrrrrrl@{}}
\hline
\multicolumn{1}{c} {Strange Stars Candidates}&  
\multicolumn{1}{c} {$M_{GR}/M_{\odot}$}& 
\multicolumn{1}{c} {$R\ [\text{km}]$}& 
 \multicolumn
 {1}{c} {$u_{GR}=M_{GR}/R$}&
 \multicolumn{1}{c}{$z_{s}$}&
 \multicolumn{1}{c}{$M/M_{\odot}$}&
 \multicolumn{1}{c}{$u=M/R$} &
 \multicolumn{1}{c}{$z_{s}$(MGD)}
 \\
\hline
$\text{Her}\ X-1$ ~\ \cite{abu} &$0.85$&$8.1$& $0.154595$&$0.203153$ &$0.87$&$0.157403$&$0.208073$ \\
\hline
$\text{SMC}\ X-1$ \cite{rawls} &$1.04$&$8.301$& $0.184571$&$0.259026$ &$1.06$&$0.187469$&$0.264848$  \\
\hline
$\text{LMC}\ X-4$ \cite{rawls} &$1.29$&$8.831$& $0.215199$&$0.324997$ &$1.31$&$0.218044$&$0.331663$  \\
\hline
\end{tabular*}
\end{table}

\begin{table}[H]
\caption{The numerical values of constant parameters $C$, $A$ and $B$ for different values of $M$ and $R$ mentioned in table \ref{table1} and $a=-1$, $b=4$ and $\alpha=-1.0$. }
\label{table2}
\begin{tabular*}{\textwidth}{@{\extracolsep{\fill}}lrrrrrrrl@{}}
\hline
\multicolumn{1}{c} {Strange Stars candidates}&  
\multicolumn{1}{c} {$C \ [\text{km}^{-2}]$}& 
\multicolumn{1}{c} {$A$\ (\text{Dimensionless})} &
\multicolumn{1}{c} {$B$\ (\text{Dimensionless})}\\
\hline
$\text{Her}\ X-1$ ~\ \cite{abu}  &$0.0009287$&$0.5408387$& $-2.4614406$ \\
\hline
$\text{SMC}\ X-1$ \cite{rawls}  &$0.0011763$&$0.4576442$& $-2.0636227$   \\
\hline
$\text{LMC}\ X-4$ \cite{rawls} &$0.0013375$&$0.3754957$& $-1.6360451$   \\
\hline
\end{tabular*}
\end{table}

\begin{table}[H]
\caption{The numerical values for central and surface density, central pressure, critical adiabatic index and central adiabatic index for different values listed in tables \ref{table1} and \ref{table2}, taking $a=-1$, $b=4$ and $\alpha=-1.0$. }
\label{table3}
\begin{tabular*}{\textwidth}{@{\extracolsep{\fill}}lrrrrrrrl@{}}
\hline
\multicolumn{1}{c} {Strange Stars}&  
\multicolumn{1}{c} {$\rho(0)$}& 
\multicolumn{1}{c} {$\rho(R)$}& 
 \multicolumn{1}{c} {$p_{r}(0)$}& \multicolumn{1}{c}{$\Gamma_{\text{crit}}$} & \multicolumn{1}{c}{$\Gamma$}  \\
 Candidates &$\times 10^{15}\  [\text{g}/\text{cm}^{3}]$&$\times 10^{14}\  [\text{g}/\text{cm}^{3}]$&$\times 10^{35}\  [\text{dyne}/\text{cm}^{2}]$\\ 
\hline
$\text{Her}\ X-1$ ~\ \cite{abu} &$0.91838$&$6.94944$& $0.83461$&$1.47575$ & $3.87767$  \\
\hline
$\text{SMC}\ X-1$ \cite{rawls}  &$1.08795$&$7.68708$& $1.28333$&$1.50295$&$3.18179$   \\
\hline
$\text{LMC}\ X-4$ \cite{rawls} &$1.14485$&$7.55726$& $1.73542$&$1.53061$&$2.67374$   \\
\hline
\end{tabular*}
\end{table}

\begin{table}[H]
\caption{The numerical values for the total mass, compactness factor and surface gravitational red--shift for $a=-1$, $b=4$ and $\alpha=-2.0$. }
\label{table4}
\begin{tabular*}{\textwidth}{@{\extracolsep{\fill}}lrrrrrrrl@{}}
\hline
\multicolumn{1}{c} {Strange Stars Candidates}&  
\multicolumn{1}{c} {$M_{GR}/M_{\odot}$}& 
\multicolumn{1}{c} {$R\ [\text{km}]$}& 
 \multicolumn
 {1}{c} {$u_{GR}=M_{GR}/R$}&
 \multicolumn{1}{c}{$z_{s}$}&
 \multicolumn{1}{c}{$M/M_{\odot}$}&
 \multicolumn{1}{c}{$u=M/R$} &
 \multicolumn{1}{c}{$z_{s}$(MGD)}
 \\
\hline
$\text{Her}\ X-1$ ~\ \cite{abu} &$0.85$&$8.1$& $0.154595$&$0.203153$ &$0.88$&$0.160209$&$0.21305$ \\
\hline
$\text{SMC}\ X-1$ \cite{rawls} &$1.04$&$8.301$& $0.184571$&$0.259026$ &$1.07$&$0.190366$&$0.27075$  \\
\hline
$\text{LMC}\ X-4$ \cite{rawls} &$1.29$&$8.831$& $0.215199$&$0.324997$ &$1.32$&$0.220889$&$0.33843$  \\
\hline
\end{tabular*}
\end{table}

\begin{table}[H]
\caption{The numerical values of constant parameters $C$, $A$ and $B$ for different values of $M$ and $R$ mentioned in table \ref{table4} and $a=-1$, $b=4$ and $\alpha=-2.0$. }
\label{table5}
\begin{tabular*}{\textwidth}{@{\extracolsep{\fill}}lrrrrrrrl@{}}
\hline
\multicolumn{1}{c} {Strange Stars candidates}&  
\multicolumn{1}{c} {$C \ [\text{km}^{-2}]$}& 
\multicolumn{1}{c} {$A$\ (\text{Dimensionless})} &
\multicolumn{1}{c} {$B$\ (\text{Dimensionless})}\\
\hline
$\text{Her}\ X-1$ ~\ \cite{abu}  &$0.0009545$&$0.5329893$& $-2.3111014$ \\
\hline
$\text{SMC}\ X-1$ \cite{rawls}  &$0.0012081$&$0.4497442$& $-1.9302915$   \\
\hline
$\text{LMC}\ X-4$ \cite{rawls} &$0.0014077$&$0.3680033$& $-1.5194916$   \\
\hline
\end{tabular*}
\end{table}

\begin{table}[H]
\caption{The numerical values for central and surface density, central pressure, critical adiabatic index and central adiabatic index for different values listed in tables \ref{table4} and \ref{table5}, taking $a=-1$, $b=4$ and $\alpha=-2.0$. }
\label{table6}
\begin{tabular*}{\textwidth}{@{\extracolsep{\fill}}lrrrrrrrl@{}}
\hline
\multicolumn{1}{c} {Strange Stars}&  
\multicolumn{1}{c} {$\rho(0)$}& 
\multicolumn{1}{c} {$\rho(R)$}& 
 \multicolumn{1}{c} {$p_{r}(0)$}& \multicolumn{1}{c}{$\Gamma_{\text{crit}}$} & \multicolumn{1}{c}{$\Gamma$}  \\
 Candidates &$\times 10^{15}\  [\text{g}/\text{cm}^{3}]$&$\times 10^{14}\  [\text{g}/\text{cm}^{3}]$&$\times 10^{35}\  [\text{dyne}/\text{cm}^{2}]$\\ 
\hline
$\text{Her}\ X-1$ ~\ \cite{abu} &$0.95929$&$6.99941$& $0.81191$&$1.47828$ & $3.07256$  \\
\hline
$\text{SMC}\ X-1$ \cite{rawls}  &$1.14006$&$7.71957$& $1.24992$&$1.50557$&$2.47052$   \\
\hline
$\text{LMC}\ X-4$ \cite{rawls} &$1.23529$&$7.69822$& $1.73586$&$1.53318$&$2.02458$   \\
\hline
\end{tabular*}
\end{table}

\begin{table}[H]
\caption{The numerical values for central and surface density, central pressure, critical adiabatic index and central adiabatic index for the GR case ($\alpha=0.0$). }
\label{table7}
\begin{tabular*}{\textwidth}{@{\extracolsep{\fill}}lrrrrrrrl@{}}
\hline
\multicolumn{1}{c} {Strange Stars}&  
\multicolumn{1}{c} {$\rho(0)$}& 
\multicolumn{1}{c} {$\rho(R)$}& 
 \multicolumn{1}{c} {$p_{r}(0)$}& \multicolumn{1}{c}{$\Gamma_{\text{crit}}$} & \multicolumn{1}{c}{$\Gamma$}  \\
 Candidates &$\times 10^{15}\  [\text{g}/\text{cm}^{3}]$&$\times 10^{14}\  [\text{g}/\text{cm}^{3}]$&$\times 10^{35}\  [\text{dyne}/\text{cm}^{2}]$\\ 
\hline
$\text{Her}\ X-1$ ~\ \cite{abu} &$0.87932$&$6.89622$& $0.85361$&$1.47321$ & $5.10185$  \\
\hline
$\text{SMC}\ X-1$ \cite{rawls}  &$1.03832$&$7.65191$& $1.31203$&$1.50033$&$4.31034$   \\
\hline
$\text{LMC}\ X-4$ \cite{rawls} &$1.11676$&$7.67022$& $1.81969$&$1.52804$&$3.76308$   \\
\hline
\end{tabular*}
\end{table}

\section{Results and Discussion}\label{section5}
In this section the most important results obtained in this work are discussed and highlighted. These results concern the obtained inner geometry describing the stellar interior, the behaviour of the principal thermodynamic observables such as density $\rho$, radial pressure $p_{r}$ and tangential pressure $p_{t}$. Besides, the energy conditions, equilibrium and stability mechanism are analyzed in order to verify if the model could represent compact structures such as neutron or quark stars.\\

We start the discussion analyzing the obtained geometry given by Eq. (\ref{eq40}). As can be seen the deformation function $f(r)$ (\ref{eq32}) is vanishing at the center of the star $r=0$ and completely regular at every point inside the structure. This fact ensures that the deformed metric potential $e^{-\lambda}\mapsto \mu(r)+\alpha f(r)$ matches with the result obtained in general relativity \i.e, $e^{\lambda(r)}|_{r=0}=1$  as expected. The trend of both metric potentials is depicted in Fig. \ref{fig1} for $\alpha=-1.0$ (left panel) and $\alpha=-2.0$ (right panel). The solid line represents the inverse of the radial metric potential $e^{-\lambda}$ while the dashed one the temporal potential $e^{\nu}$.
As illustrates Fig. \ref{fig1} the temporal component of the interior space--time respects the pertinent requirements that is, monotone increasing function with increasing radial coordinate and $e^{\nu(0)}>0$ at $r=0$. Besides, as it is observed both metric potentials coincide at the boundary of the compact structure. This shows that the junction condition process is correct. 

\begin{figure}[H]
\centering
\includegraphics[width=0.32\textwidth]{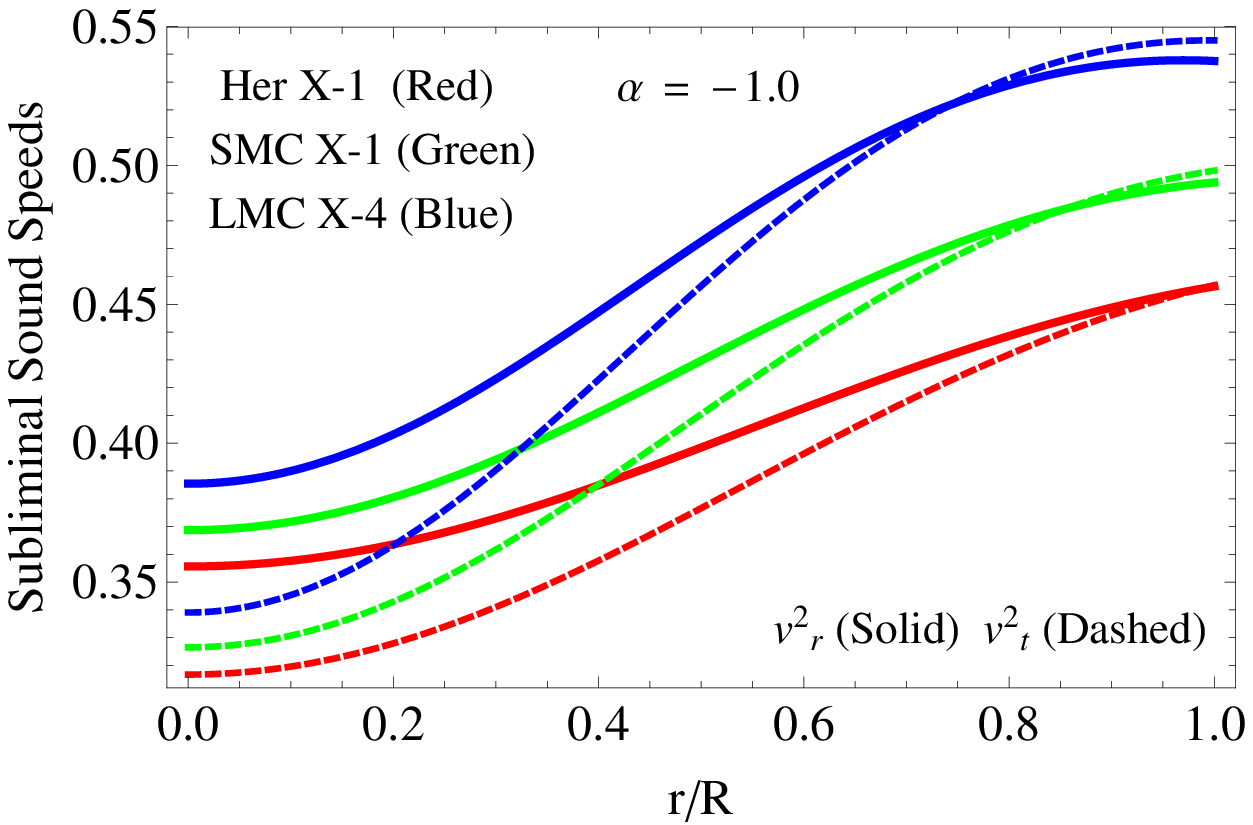}     
\includegraphics[width=0.32\textwidth]{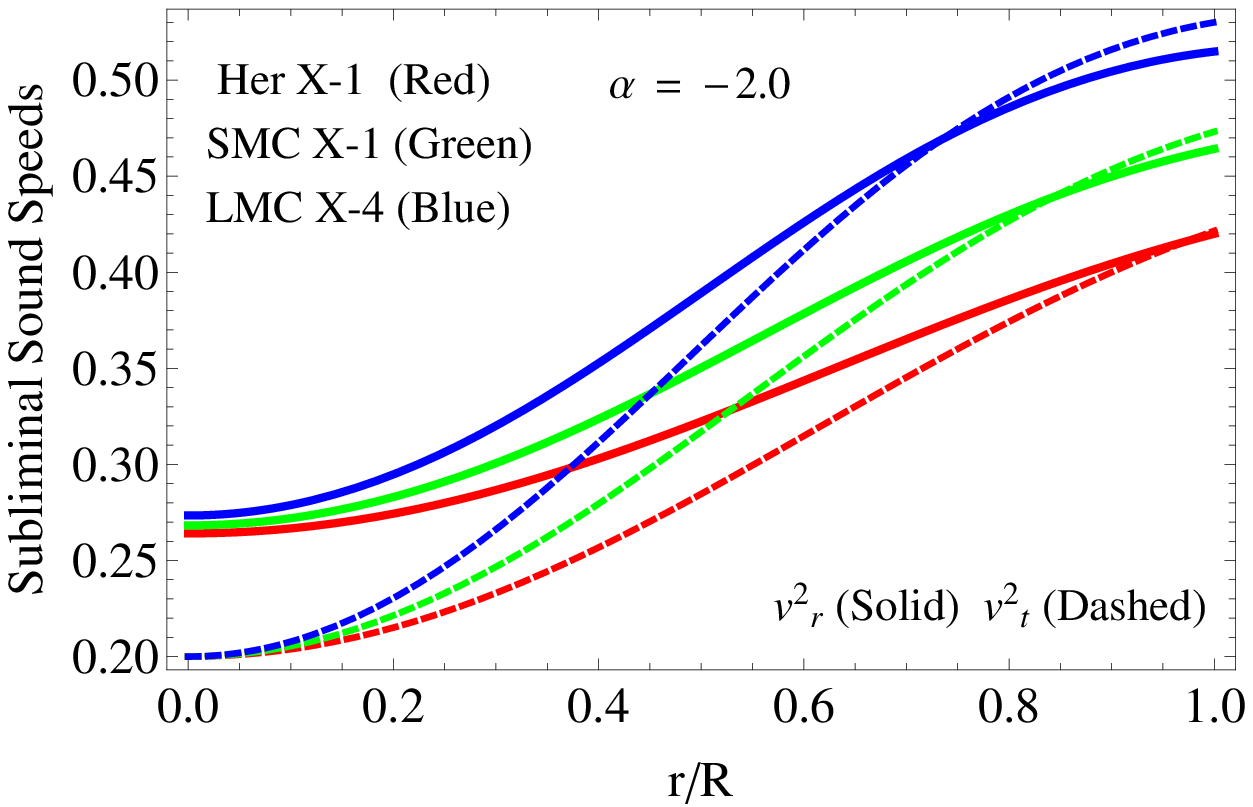}      \\
\includegraphics[width=0.34\textwidth]{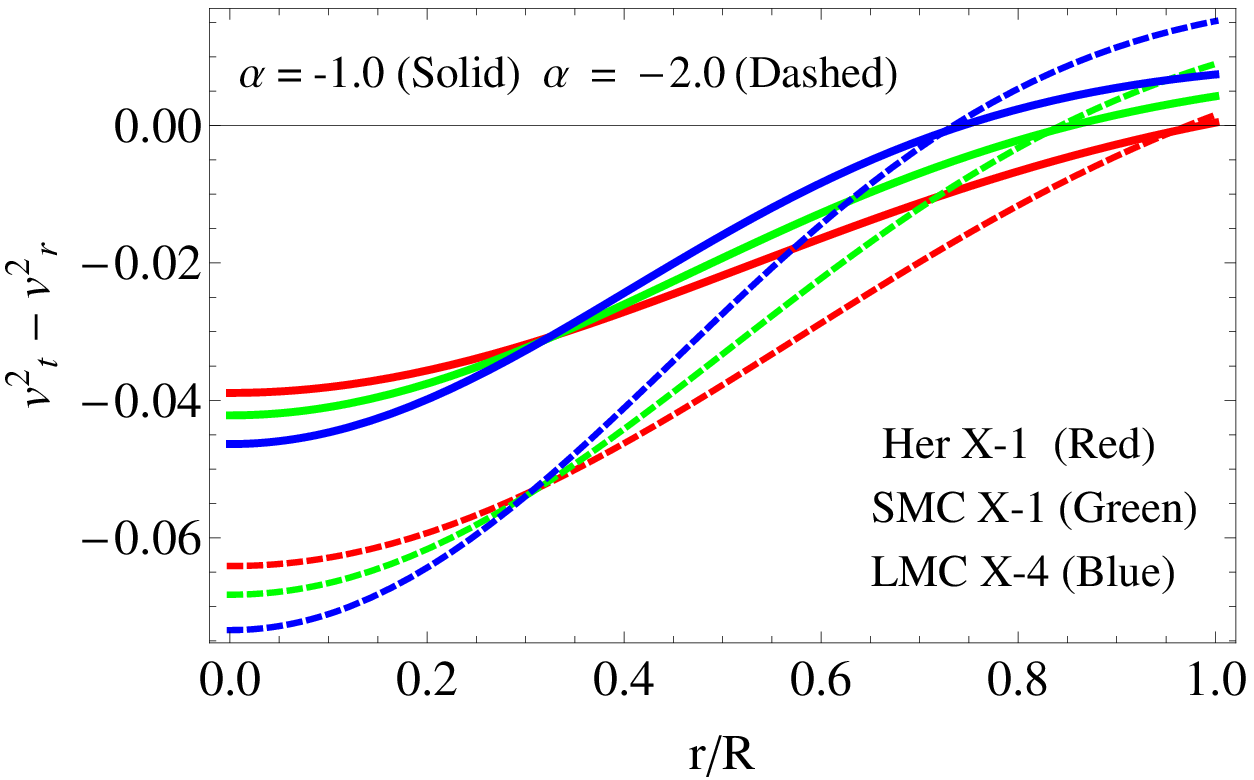} 
\includegraphics[width=0.32\textwidth]{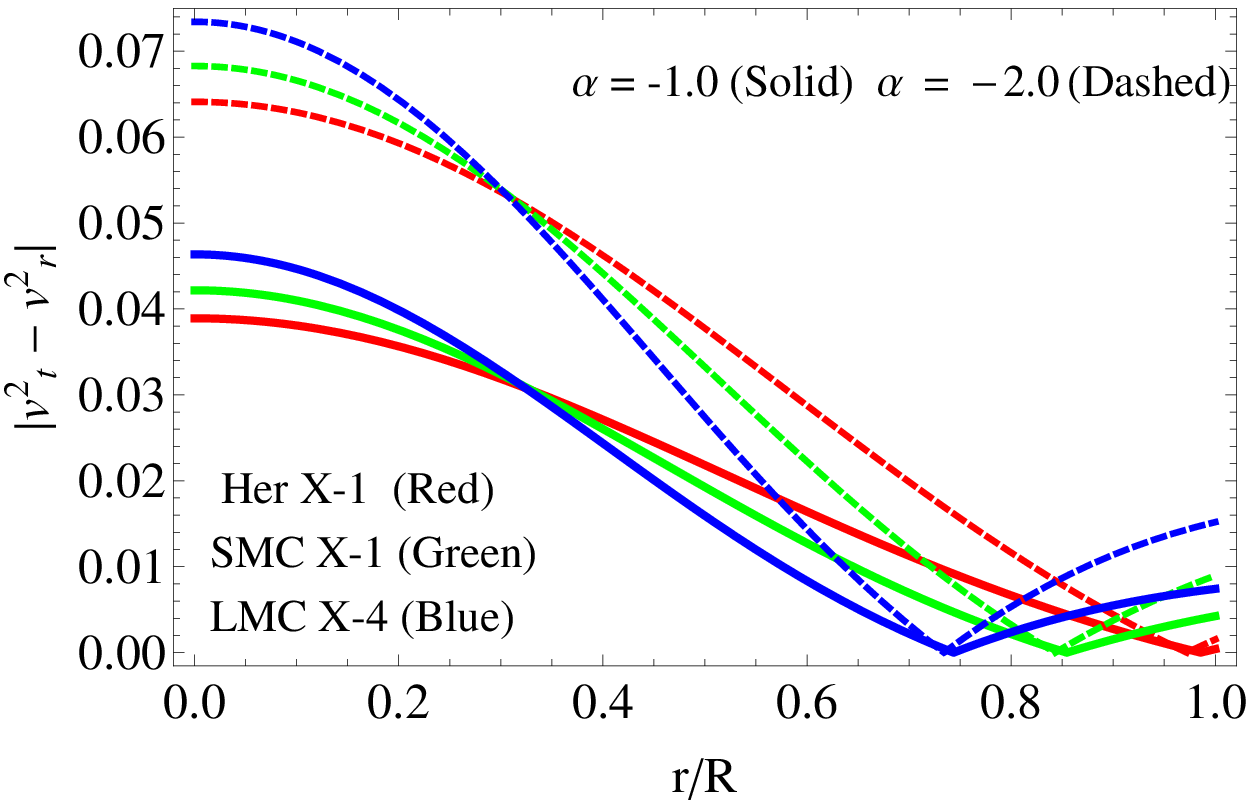} 
\caption{
{\bf Upper row}: The square of sound velocity of pressure waves in the principal directions for $\alpha=-1.0$ (left side) and $\alpha=-2.0$ (right side), respectively. {\bf Lower left panel}: The different of the squared sound velocities against the dimensionless radial coordinate. {\bf Lower right panel}: Abreu's factor versus the radial coordinate $r/R$ 
}
\label{fig3}
\end{figure}

The principal thermodynamic variables driving the matter distribution inside the fluid sphere are shown in Fig. \ref{fig2}. As can be seen in Fig. \ref{fig2} (Upper panels) the radial and tangential pressures coincide at the center of the structure and drift apart towards the boundary. Moreover, the transverse pressure dominates the radial one in all cases. It is worth mentioning that the central pressure increases with increasing mass. Furthermore, from Fig. \ref{fig2} (Lower left panel) it is clear that the anisotropy quantifier $\Delta$ is a monotone increasing function with increasing radius. Of course, as can be seen at the center of the object we have $\Delta=0$ due to at $r=0$ both pressures are equal. However, at the surface of the star the blue curve takes greater values in comparison with the red and green curves. This means that the difference between $p_{r}$ and $p_{t}$ increases as the total mass increases. At this point it should be noted that in order to have a positive anisotropy factor $\Delta$ inside the stellar interior from Eq. (\ref{eq39}) is evident that $\alpha$ must be negative. As pointed out by Gokhroo and Mehra a positive anisotropy factor allows the construction of more compact objects \cite{r13}. With respect to the energy--density $\rho$ from the lower right panel of Fig. \ref{fig2} it is observed that in all cases this observable takes its upper value at the center of the star and decreases monotonically towards the boundary $\Sigma$. An interesting point to be highlighted here, is that the central parameters $\rho(0)$ and $p_{r}(0)$ and the surface density $\rho(R)$ increase with increasing mass and increasing $\alpha$ magnitude (for further details see tables \ref{table3} and \ref{table6}). On the other hand, as shown table \ref{table7} the central and surface density in the GR case are dominated when MGD is present, while GR central pressure dominates the MGD scenario.
Next, from the upper left ($\alpha=-1.0$) and right ($\alpha=-2.0$) panels of Fig. \ref{fig3} is illustrated that the velocities of the pressure waves in the main directions, expressed by
\begin{equation}
v^{2}_{r}(r)=\frac{dp_{r}(r)}{d\rho (r)} \quad \mbox{and}\quad v^{2}_{t}(r)=\frac{dp_{t}(r)}{d\rho (r)},  
\end{equation}
are less than the unit (where $c=1$), then the present model satisfies causality condition. Nevertheless, as the mass increases occurs a swap between $v_{r}$ and $v_{t}$. This fact is related with the instabilities against cracking process \cite{r12} introduced by the anisotropies in the stellar matter distribution. To check the unstable regions inside the star we have plotted the so called Abreu's factor $|v^{2}_{t}-v^{2}_{r}|$ \cite{r20} (lower right panel in Fig. \ref{fig3}) and the difference of the square sound velocities (lower left panel in Fig. \ref{fig3}). The change in sign of $v^{2}_{t}-v^{2}_{r}$ means that the system present unstable regions \i.e, zones with cracking. 
Explicitly, the stable/unstable regions within the stellar interior, when local anisotropies are there can be found as \cite{r20},
\begin{equation}\label{eq54}
\frac{\delta{\Delta}}{\delta{\rho}}\sim \frac{\delta\left({p}_{t}-{p}_{r}\right)}{\delta{\rho}} \sim \frac{\delta{p}_{t}}{\delta{\rho}}-\frac{\delta{p}_{r}}{\delta{\rho}}\sim v^{2}_{t}-v^{2}_r.    
\end{equation}
Taking into account (\ref{eq54}), one gets  $0\leq |v^{2}_{t}-v^{2}_{r}|\leq 1$, which can be re--expressed as follows
\begin{equation}
\begin{split}
     \label{eq55}
   -1\leq v^{2}_{t}-v^{2}_{r}\leq 1  = \left\{
	       \begin{array}{ll}
		   -1\leq v^{2}_{t}-v^{2}_{r}\leq 0~~ & \mathrm{Potentially\ stable\ }  \\
		 0< v^{2}_{t}-v^{2}_{r}\leq 1 ~~ & \mathrm{Potentially\ unstable}
	       \end{array}
	        \right\}.
	        \end{split}
	    \end{equation}
Therefore, the compact structure will be stable under radial perturbations induced by local anisotropies, if the subliminal radial sound speed, $v^{2}_{r}$, of the pressure waves, dominates everywhere against the subliminal sound velocity of the pressure waves in the tangential direction $v^{2}_{t}$. 
As can be seen from the lower panels in Fig. \ref{fig3}, there is a clear tendency of the system to present unstable regions when the mass of the system increases. 
Also, the situation is even worse when the parameter $\alpha$ increases (in magnitude). On the other hand, for values corresponding to the star Her X--1 (the least massive star), the system does not present unstable zones. In this concern, it is clear that the mass of the object directly influences the fulfillment of this stability criterion as well as does the parameter $\alpha$. Thus, in light of the features above, one possibility to circumvent cracked areas in the stellar interior, without considering less massive objects, would be limiting the value (magnitude) of the constant coupling $\alpha$. This can also be done without loss of generality. The latter is true because $\alpha$ is a free parameter that only controls the strength of the anisotropy induced via MGD by gravitational decoupling.

\begin{figure}[H]
\includegraphics[width=0.32\textwidth]{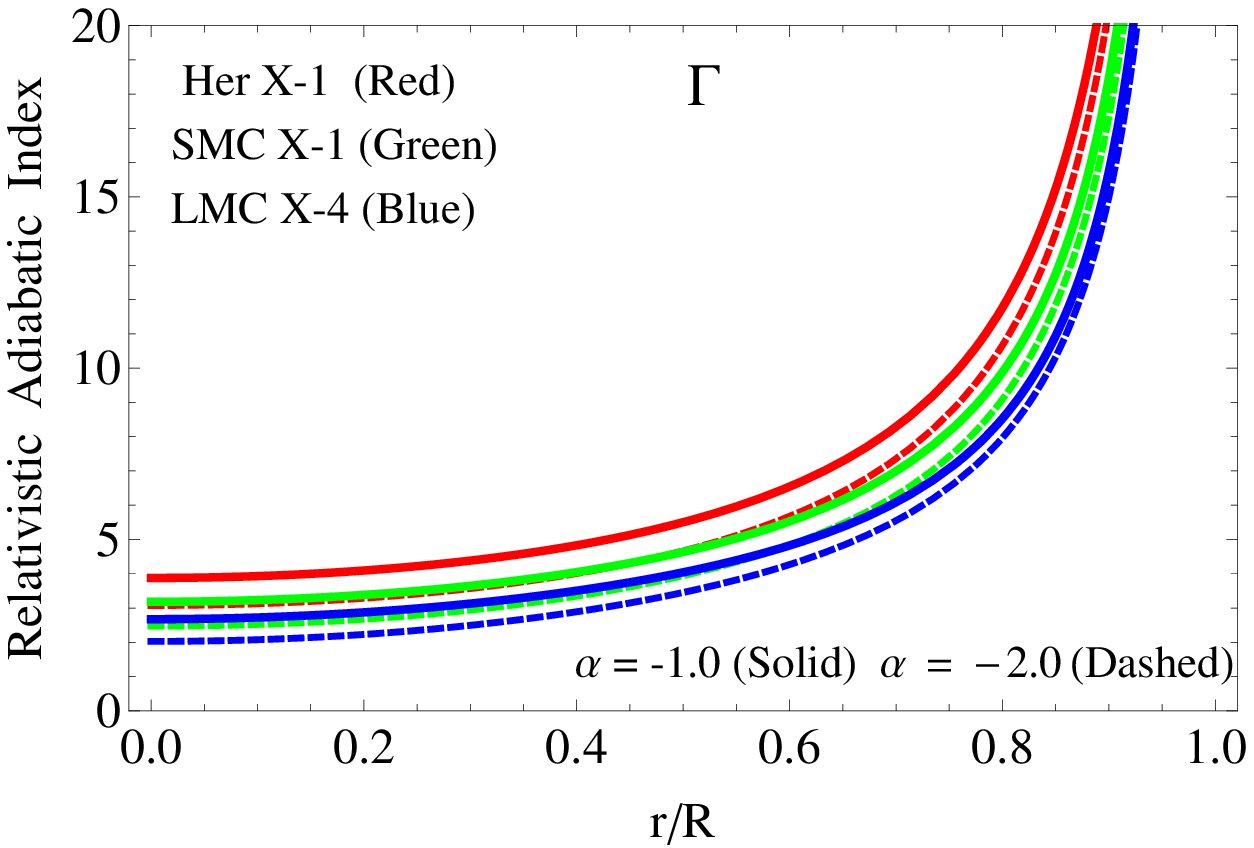} \
\includegraphics[width=0.32\textwidth]{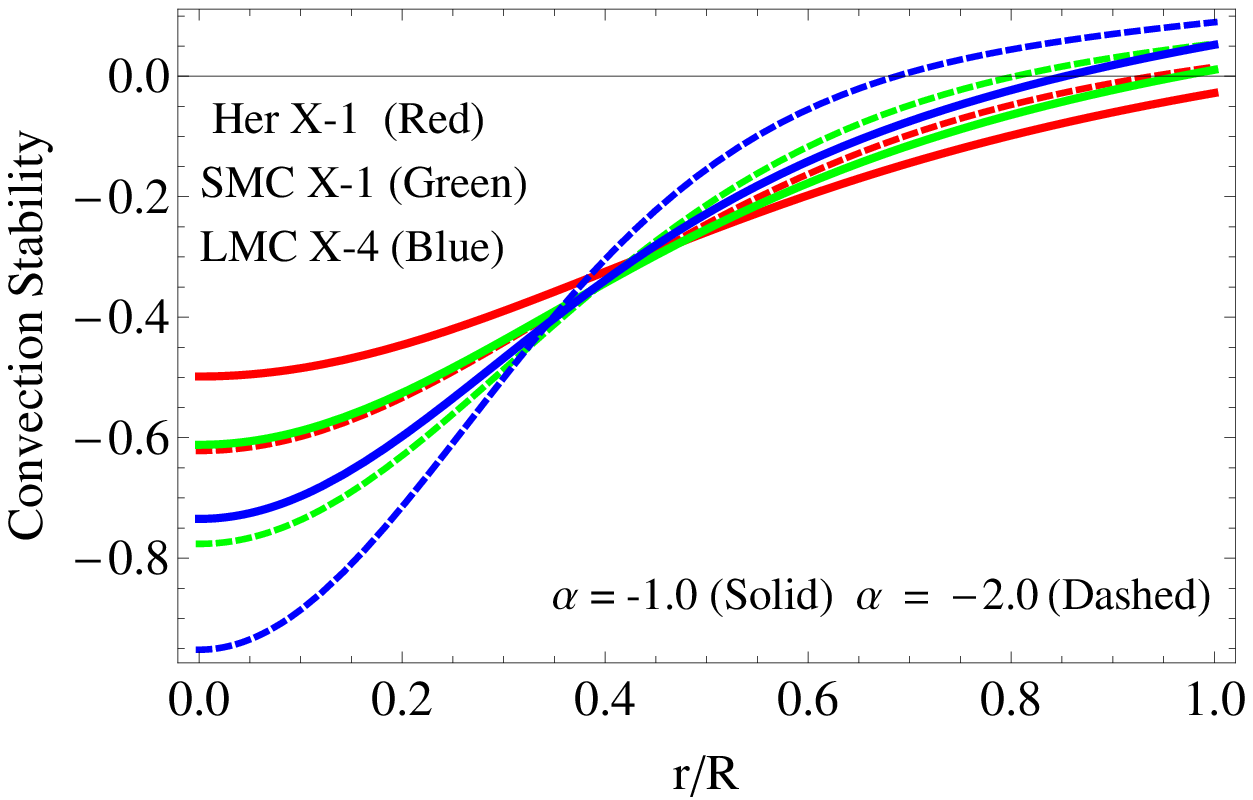}\ 
\includegraphics[width=0.32\textwidth]{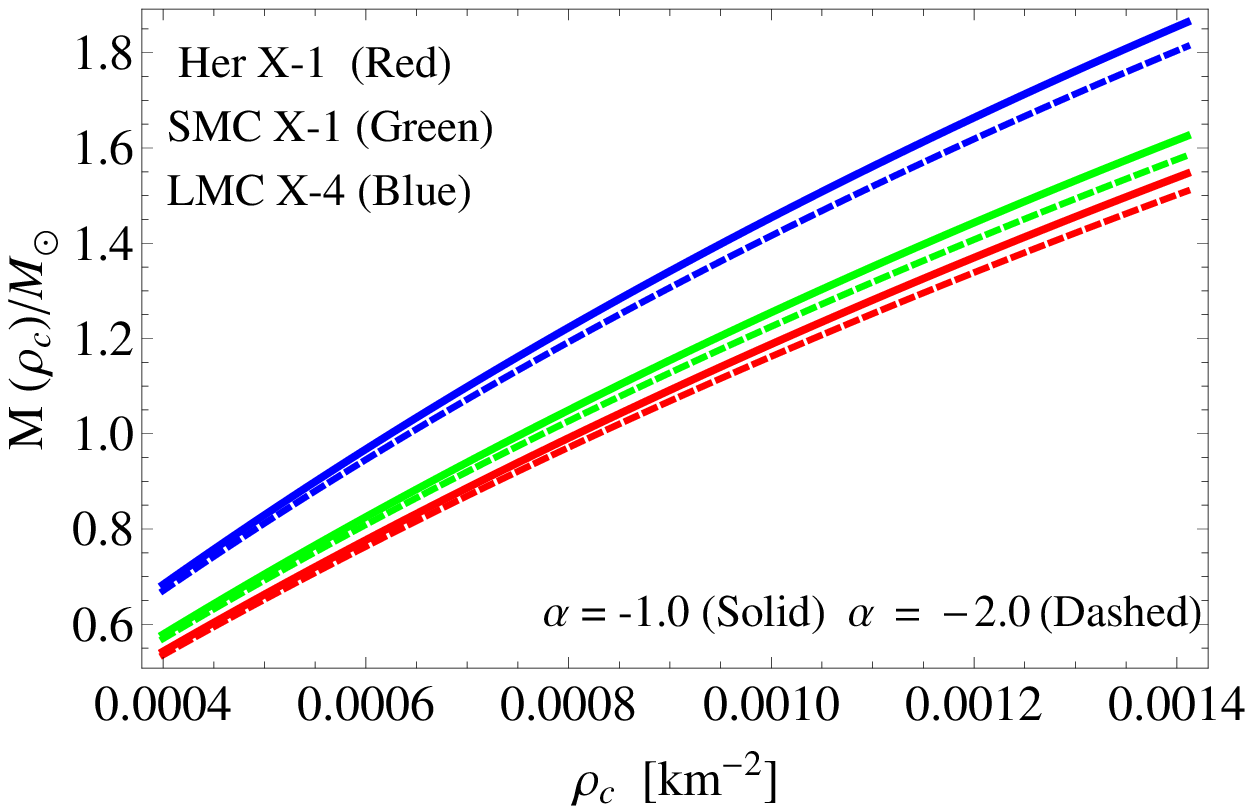}
\caption{
{\bf Left panel}: Variation of the adiabatic index $\Gamma$ versus $r/R$. {\bf Middle panel}: The convection stability factor against the dimensionless radius $r/R$. {\bf Right panel}: The variation of the total mass against the central density. }
\label{fig4}
\end{figure}

On the other hand we also have checked the stability of the configuration using the relativistic adiabatic index $\Gamma$. For a Newtonian isotropic fluid distribution the stability condition is $\Gamma>4/3$ \cite{r1,r9}. However, in the anisotropic relativistic scenario the above condition is quite different. In that case the system should satisfies \cite{r10,r11}
\begin{equation}\label{adibatic}
\Gamma>\frac{4}{3}+\left[\frac{1}{3}\kappa\frac{\rho_{0}p_{r0}}{|p^{\prime}_{r0}|}r+\frac{4}{3}\frac{\left(p_{t0}-p_{r0}\right)}{|p^{\prime}_{r0}|r}\right]_{max}    
\end{equation}
where $\rho_{0}$ corresponds to the initial density and $p_{r0}$ represents the radial pressure (when the fluid is in static equilibrium). Finally, $p_{t0}$  is the corresponding tangential pressure (again when the fluid is in static equilibrium). 
The second term in the right hand side represents the relativistic corrections to the Newtonian perfect fluid and the third term is the contribution due to anisotropy. 
It is observed that the stability condition is modified due to  relativistic corrections and the presence of local anisotropies. However, as was pointed out by Chandrasekhar \cite{chandra1,chandra2} relativistic correction to the adiabatic index could in principle introduce some instabilities within the stellar interior. This would happen in the case where the object contracts until it reaches the critical radius
\begin{equation}
R_{\text{crit}}=\frac{2KM}{\Gamma+\frac{4}{3}},    
\end{equation}
being $K$ a constant depending on the density distribution. Of course, in this case the local anisotropies help to avoid this situation. Moreover, recently a more strict condition on adiabatic index for stable region was computed in \cite{mousta}. It was found that critical value of adiabatic
index $\Gamma_{\text{crit}}$ (to have a stable structure) depends on the amplitude of the Lagrangian
displacement from equilibrium and the compactness factor $u\equiv M/R$. The amplitude of the Lagrangian displacement is characterized by the parameter $\xi$, so taking particular form of this parameter the critical relativistic adiabatic index is given by
\begin{equation}
\Gamma_{\text{crit}}=\frac{4}{3}+\frac{19}{21}u,    
\end{equation}
where the stability condition becomes  $\Gamma\geq \Gamma_{\text{crit}}$.
On the other hand the explicit form to compute the adiabatic index is given by \cite{r8}
\begin{equation}
\Gamma=\frac{\rho+p_{r}}{p_{r}}\frac{dp_{r}}{d\rho}.    
\end{equation}
As the left panel of Fig. \ref{fig4} shows the system is completely stable from the relativistic adiabatic index point of view for all cases. Besides, as the numerical values displayed in tables \ref{table3} ($\alpha=-1.0$) and \ref{table6} ($\alpha=-2.0$) corroborate the condition $\Gamma\geq \Gamma_{\text{crit}}$ is satisfied in all cases. It should be noted that as the mass of the object and the magnitude of $\alpha$ increase, the central value of $\Gamma$ decreases. Furthermore, as table \ref{table7} shows, when $\alpha=0.0$ (GR case), the adiabatic index takes higher values at $r=0$, which indicates as Chandrasekhar argued that the presence of anisotropy does not always improve the stability conditions of the system. However, despite this, the system is stable under this criterion. In a more widely context, talk about stability mechanisms is rather heuristic, in this concern there are many ways to check or analyzed the stability of a system under radial perturbations caused by the presence of anisotropies. Recently in \cite{hh} the convection stability concept was introduced. In short, a stellar interior is stable against convection when a fluid element displaced downward floats back to its initial position. This occurs whenever $\rho^{\prime\prime}<0$. In Fig. \ref{fig4} (middle panel) it is shown that the model is unstable after under going convection motion. As in the relativistic adiabatic index case, there is a clear tendency of the system to reduce the stability when both $M$ and $\alpha$ increase in magnitude. As can be appreciated, for the less massive star (Her X--1) and $\alpha=-1.0$ the system is stable under this approach, while the other cases are not. Furthermore, the same situation occurs when Abreu's criterion analysis is employed \i.e, the system seems to become unstable as the mass increases. On the other hand, a more reliable and well posed stability criteria is based on the well--known Harrison--Zeldovich–-Novikov \cite{harrison,nikolov} procedure. This criteria says that any fluid configuration is stable if the mass is a increasing function with respect to the central density $\rho(0)=\rho_{c}$ \i.e, $\frac{\partial M(\rho_{c})}{\partial \rho_{c}}>0$, otherwise the model is unstable. As the right panel in Fig. \ref{fig4} depicts, the total mass as a function of the central density is an increasing quantity. Therefore, the present model is completely stable.   

\begin{figure}[H]
\centering
\includegraphics[width=0.32\textwidth]{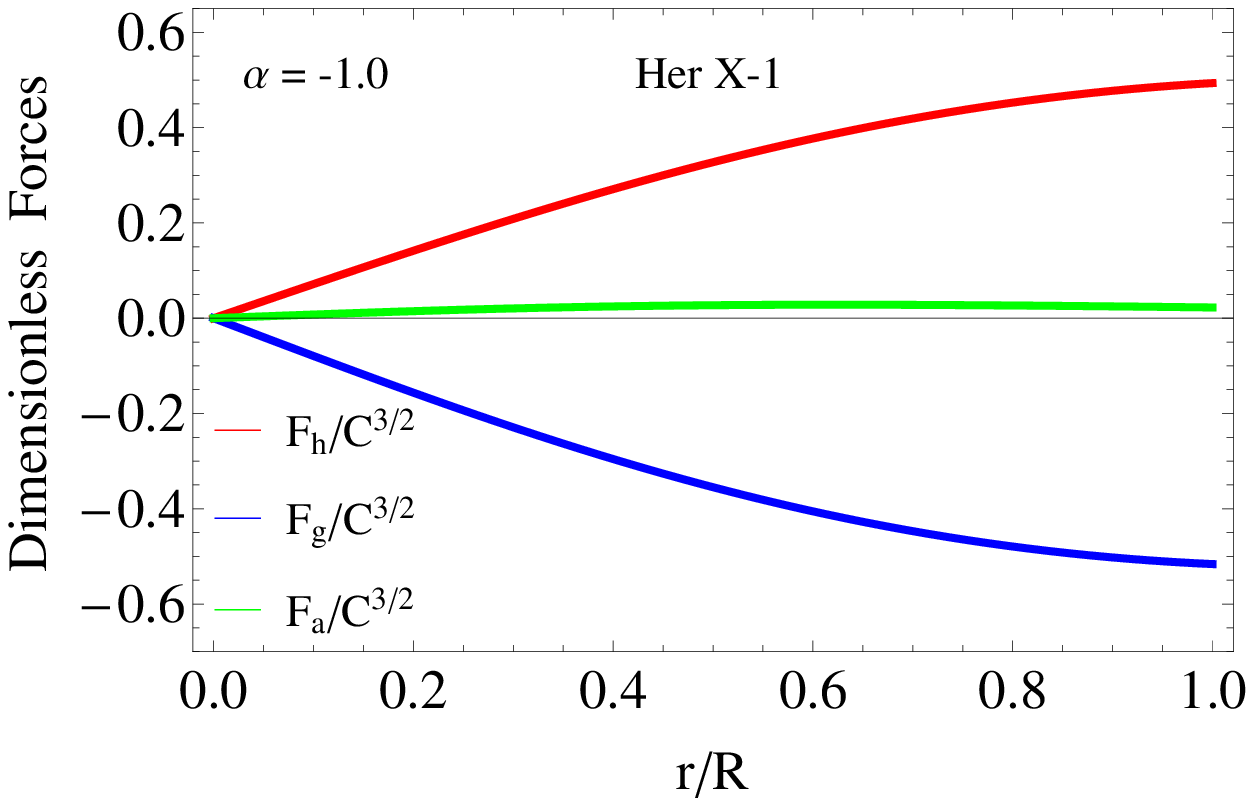}     
\includegraphics[width=0.32\textwidth]{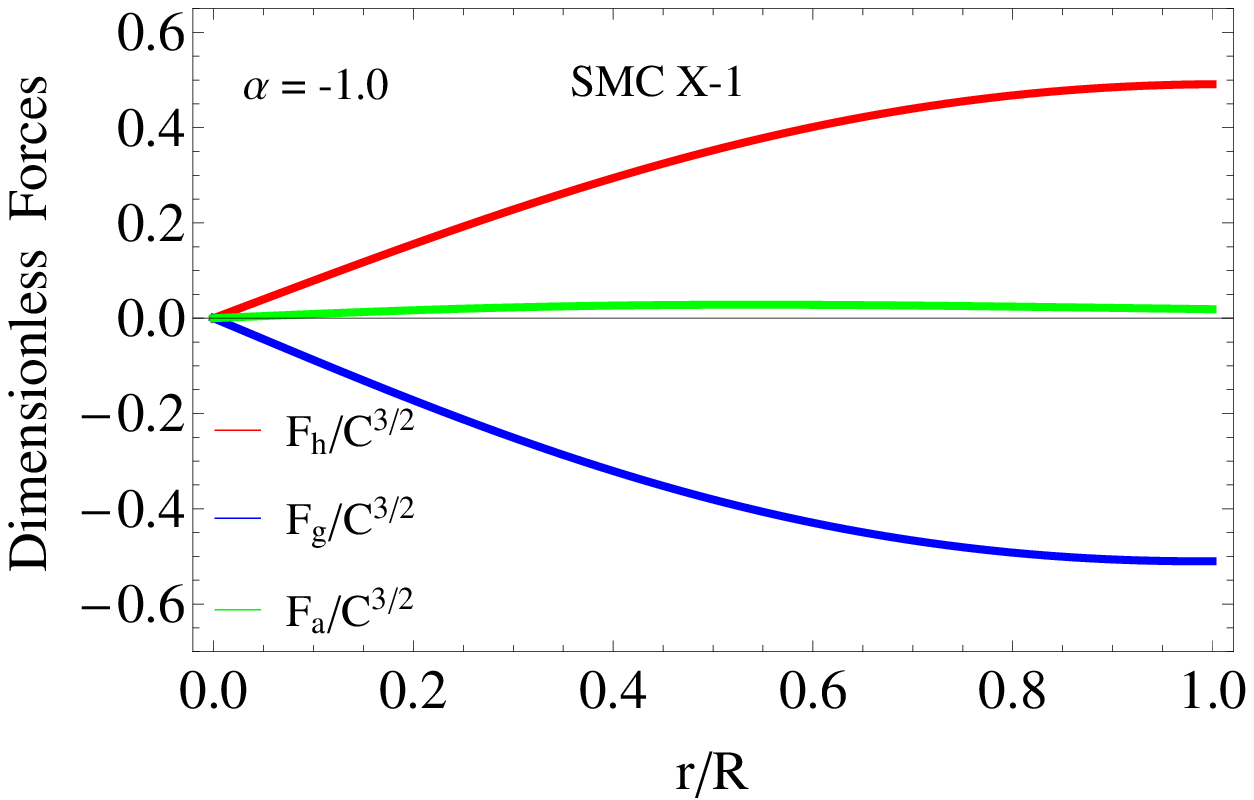}      
\includegraphics[width=0.32\textwidth]{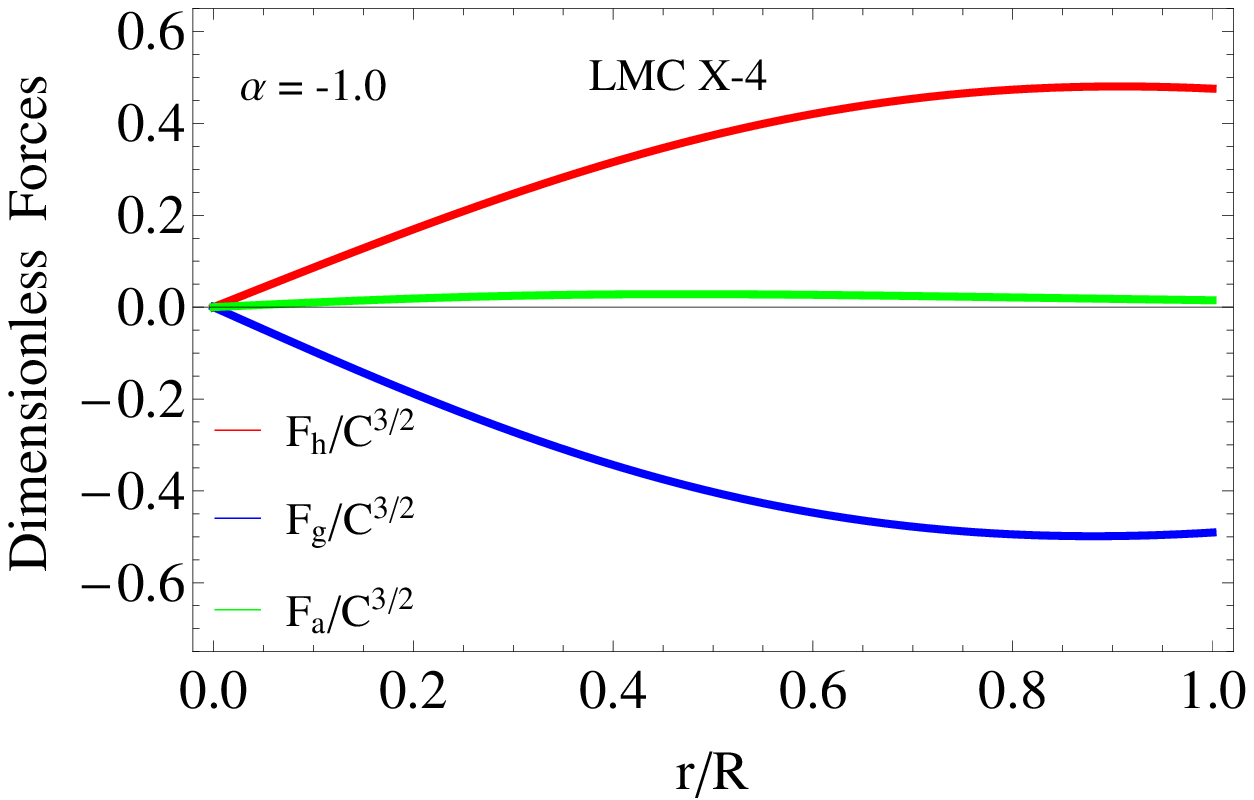}     \\
\includegraphics[width=0.32\textwidth]{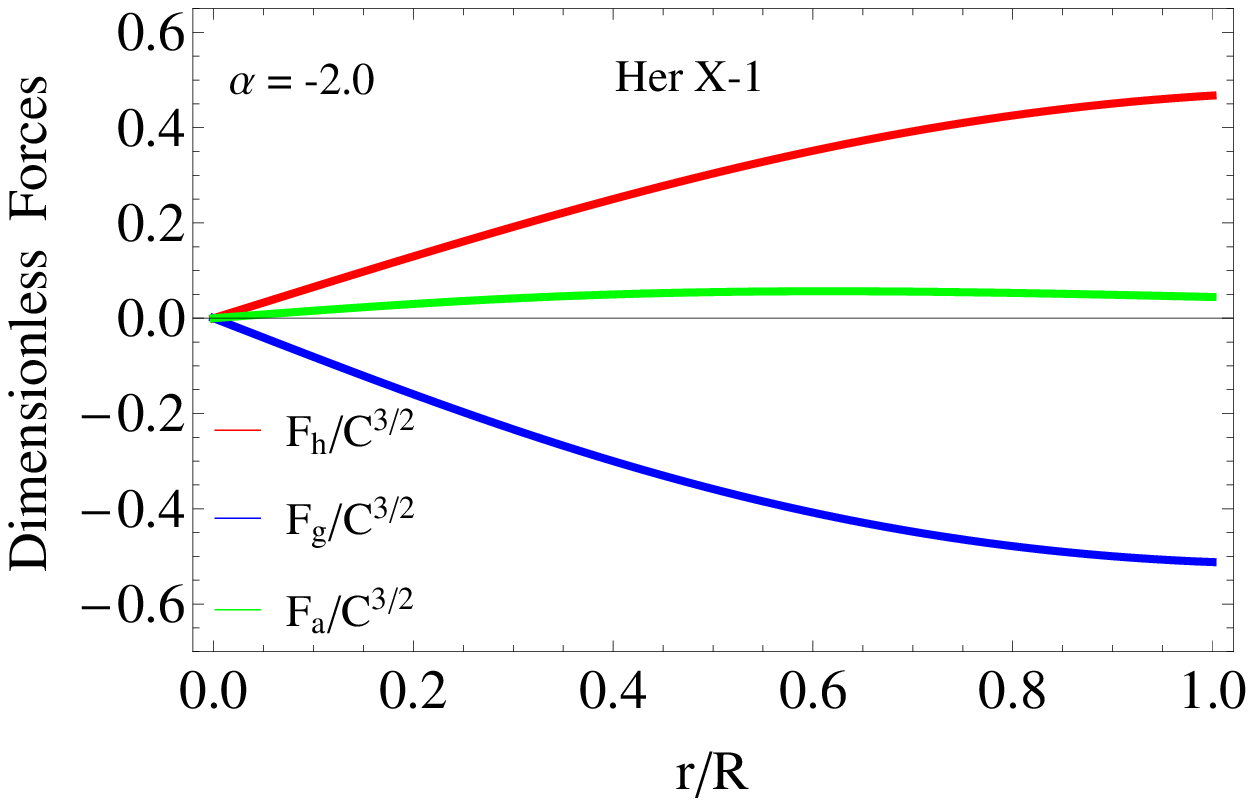}     
\includegraphics[width=0.32\textwidth]{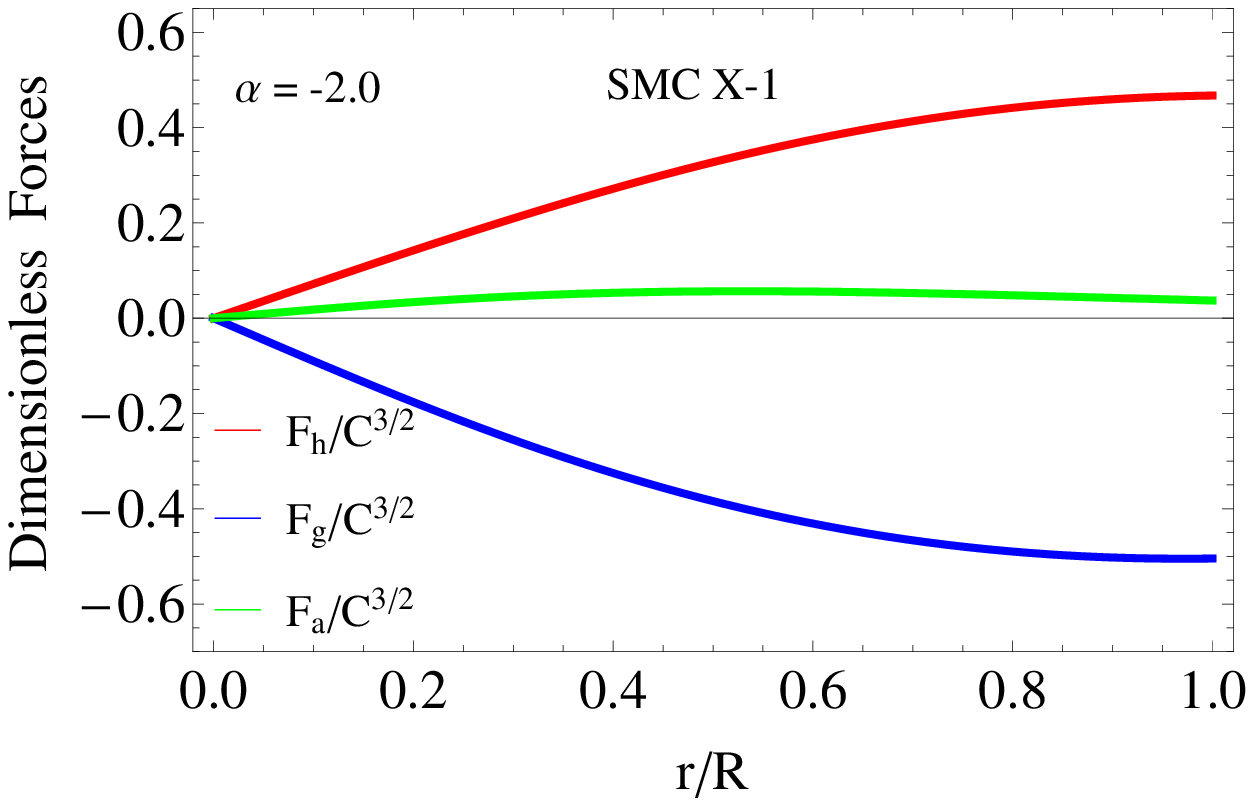}      
\includegraphics[width=0.32\textwidth]{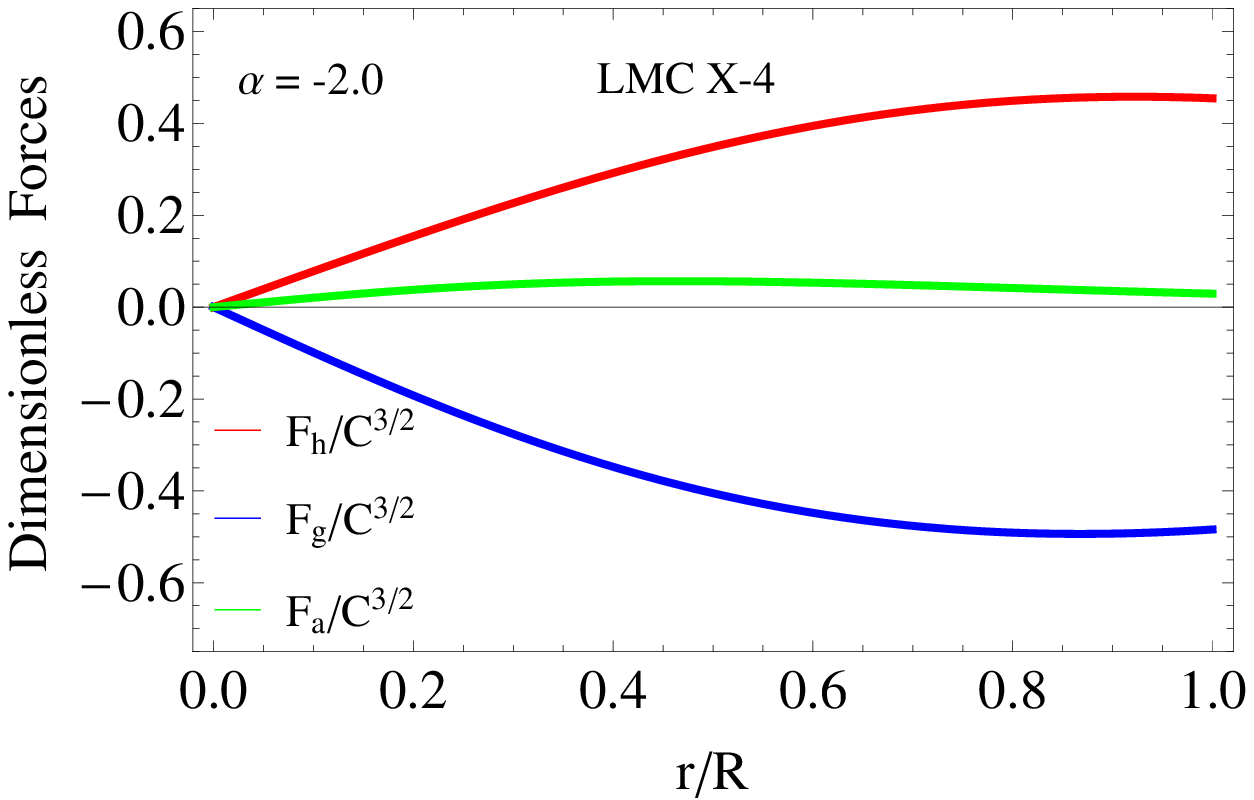} 
\caption{
The Tolman-Oppenheimer-Volkoff equation versus $r/R$, representing the balance of the system under the gravitational, hydrostatic and anisotropic forces.
}
\label{fig5}
\end{figure}

The balance of the system under different forces such as gravitational force $F_{g}$, hydrostatic force $F_{h}$ and anisotropic force $F_{a}$ is studied by employing the conservation law of the energy-momentum tensor $\nabla_{\mu}T^{\mu\nu}=0$ which yields to \cite{r46}  
\begin{equation}\label{tov}
\begin{split}
-\frac{d\tilde{p}}{dr}-\alpha\left[ \frac{\nu^{\prime}}{2}\left(\theta^{t}_{t}-\theta^{r}_{r}\right)-\frac{d \theta^{r}_{r}}{dr}+\frac{2}{r}\,(\theta^{\varphi}_{\varphi}-\theta^{r}_{r})\right]-\frac{\nu^\prime}{2} (\tilde{\rho}+\tilde{p})=0.
\end{split}
\end{equation}
The expression (\ref{tov}) can be seen as a generalization of the well known Tolman--Oppenheimer--Volkoff (TOV)  \cite{r82,r81} which describes the hydrostatic equilibrium of isotopic fluid distributions. Of course, in this case we are dealing with an anisotropic fluid sphere where the corresponding modifications are encode into the $\theta$-sector. A closer look at the equation (\ref{tov}) reveals that the hydrostatic, gravitational and anisotropy gradients are given by
\begin{eqnarray}
F_{h}&=&\frac{d}{dr}\left[\alpha\theta^{r}_{r}-\tilde{p}_{r}\right], \\ 
F_{g}&=&\frac{\nu^{\prime}}{2}\left[\alpha\left(\theta^{r}_{r}-\theta^{t}_{t}\right)-\tilde{\rho}-\tilde{p}_{r}\right], \\
F_{a}&=&\frac{2\alpha}{r}\left[\theta^{r}_{r}-\theta^{\varphi}_{\varphi}\right].
\end{eqnarray}
It is observed that gravitational decoupling by MGD modifies the usual gradients \i.e, the hydrostatic and gravitational ones, introducing and extra piece moderated by the dimensionless parameter $\alpha$. This new piece $F_{a}$ plays the role of anisotropic gradient, where if $F_{a}>0$ the system experiences a repulsive force counteracting with the help of $F_{h}$ the gravitational force $F_{g}$ to avoid the gravitational collapse of the structure onto a point singularity. It should be noted that in the limit $\alpha\rightarrow0$ the TOV equation is recovered. As Fig. \ref{fig5} illustrates the system is in equilibrium under the aforementioned forces. Notwithstanding, notice that the anisotropic force is negligible in all cases in comparison with hydrostatic and gravitational forces. Despite its small contribution the anisotropic force helps to counteract the gravitational gradient allowing to the system remains balanced against gravitational collapse.  

\begin{figure}[H]
\centering
\includegraphics[width=0.32\textwidth]{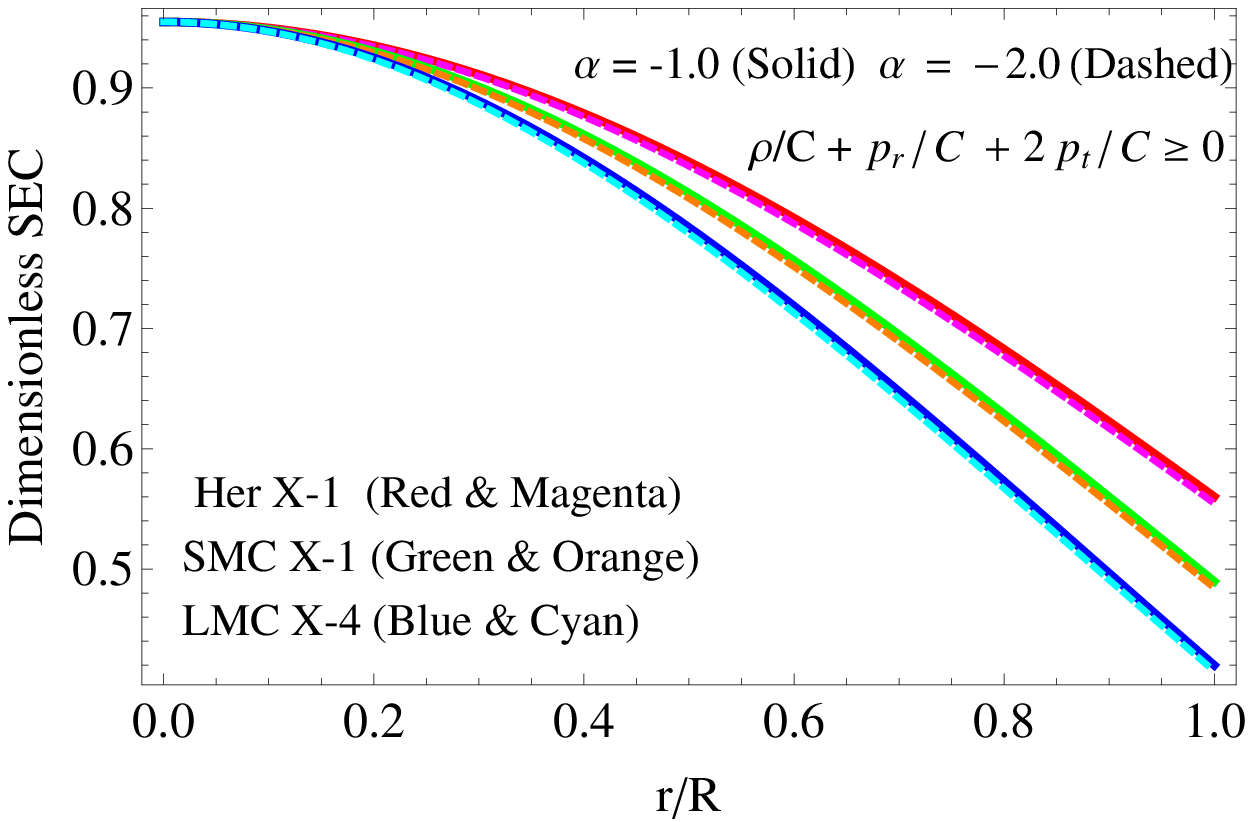} 
\includegraphics[width=0.32\textwidth]{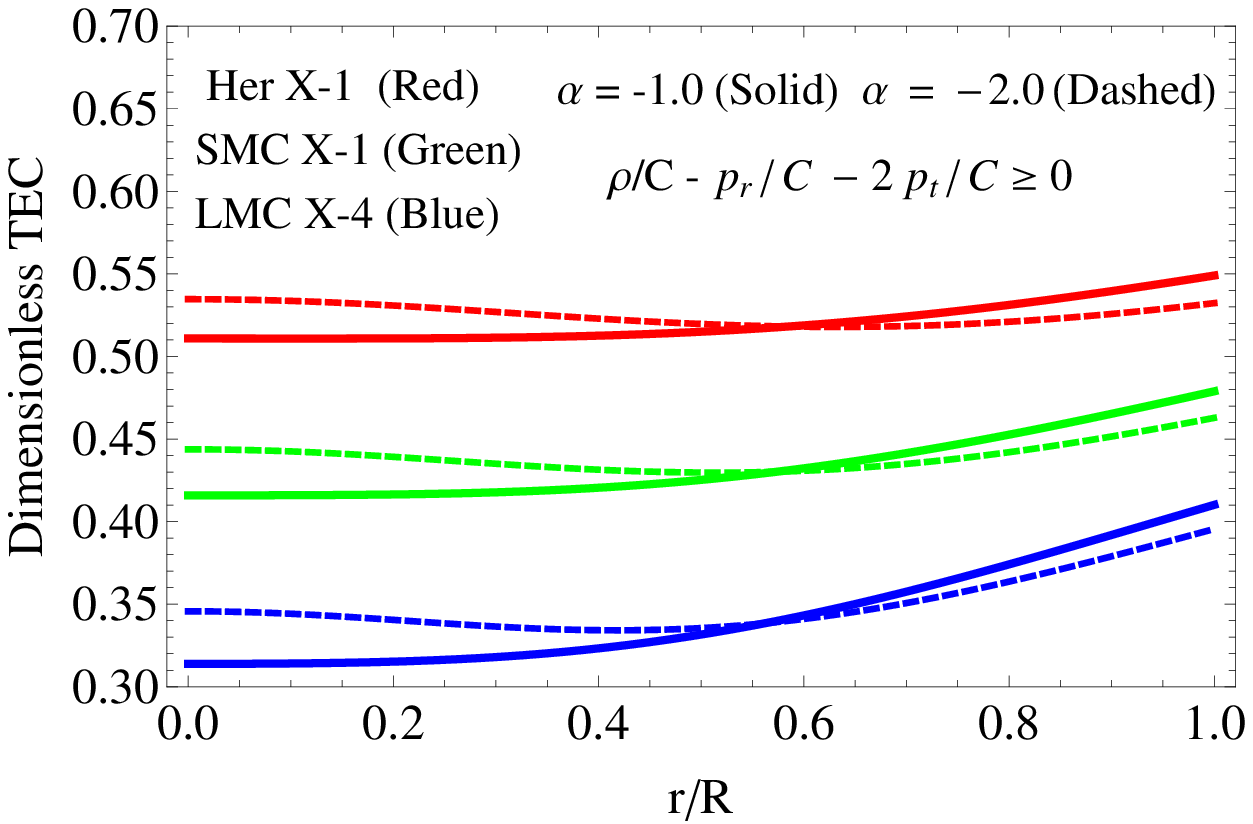}\\ 
\includegraphics[width=0.32\textwidth]{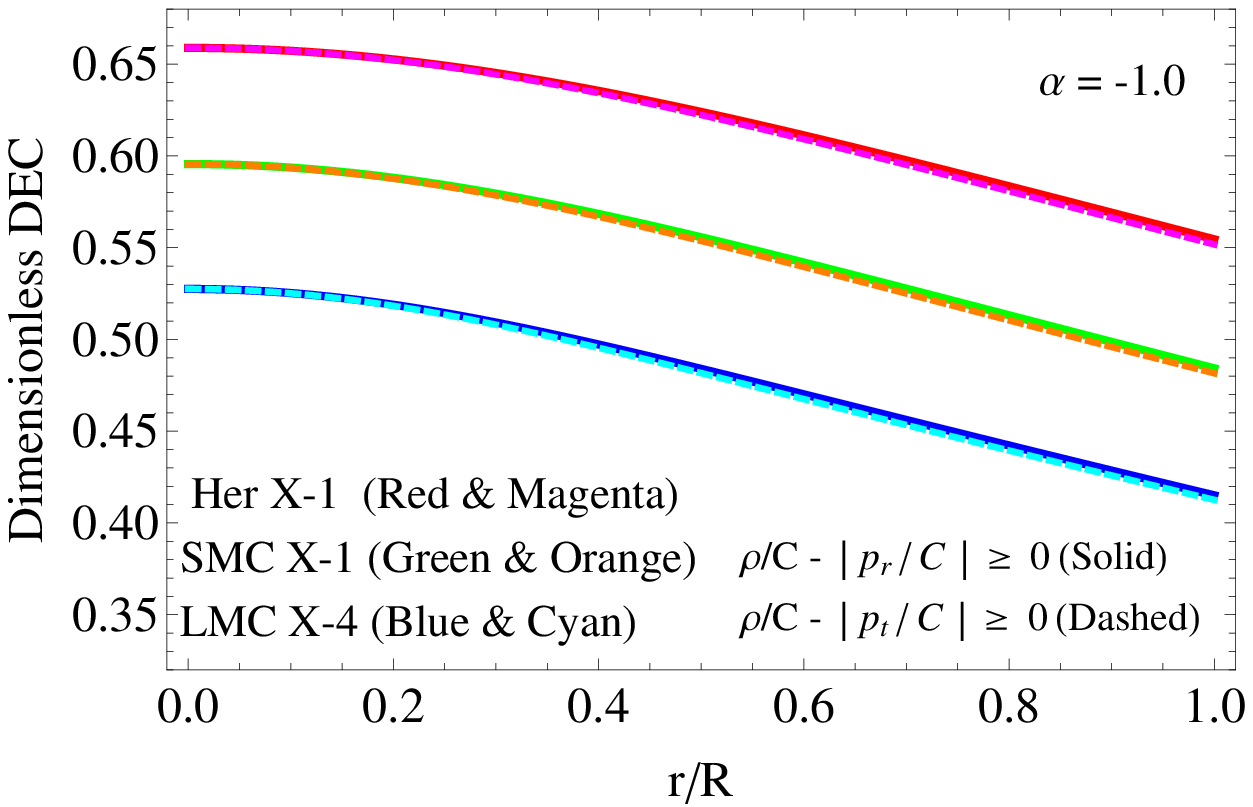} 
\includegraphics[width=0.32\textwidth]{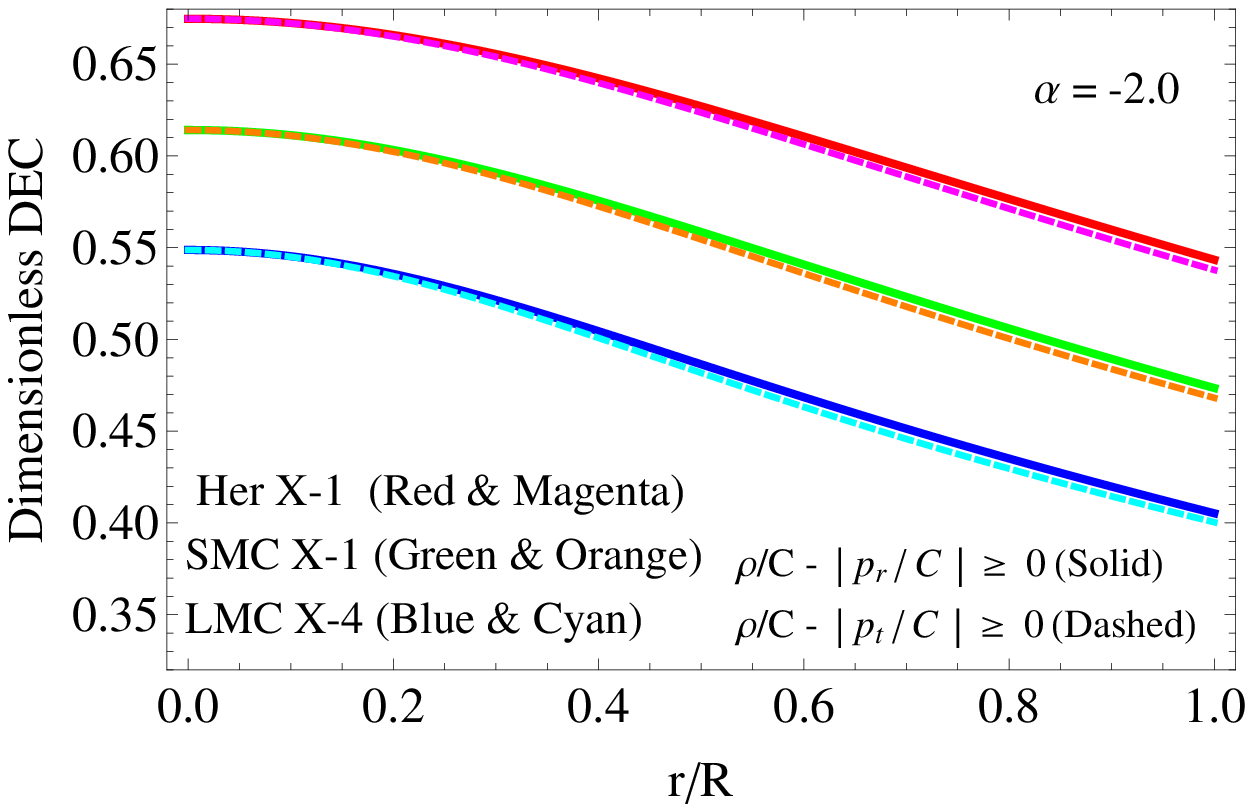}
\caption{
{\bf Upper row}: The strong energy condition (SEC) and trace energy condition (TEC) versus the dimensionless distant $r/R$.  {\bf Lower row}: The dominant energy condition (DEC) for different values exhibited in tables \ref{table2} and \ref{table5}.
}
\label{fig6}
\end{figure}

In order to verify if the energy-momentum tensor threading the material content within the star is well defined,  there are some conditions that are suitable to satisfy in order to obtain a well-defined physical system. Those are refereed as energy conditions, specifically they are: 
i) the weak energy condition (WEC), 
ii) the null energy condition (NEC), 
iii) the dominant energy condition (DEC),
iv) the strong energy condition (SEC) and finally the trace energy condition (TEC).
Explicitly, these are given by \cite{visserbook}
\begin{eqnarray}\label{eq60}
\text{WEC} &:& T_{\mu \nu}t^\mu t^\nu \ge 0~\mbox{or}~\rho \geq  0,~\rho+p_i \ge 0, \\
\text{NEC} &:& T_{\mu \nu}l^\mu l^\nu \ge 0~\mbox{or}~ \rho+p_i \geq  0,\\
\text{DEC} &:& T_{\mu \nu}t^\mu t^\nu \ge 0 ~\mbox{or}~ \rho \ge |p_i|,\\
&& \mbox{where}~~T^{\mu \nu}t_\mu \in \mbox{nonspace-like vector} \nonumber \\ 
\text{SEC} &:& T_{\mu \nu}t^\mu t^\nu - \frac{1}{2} T^\lambda_\lambda t^\sigma t_\sigma \geq 0 ~\mbox{or}~ \rho+\sum_i p_i \ge 0, \\ \label{eq63}
\text{TEC} &:& g^{\mu\nu}T_{\mu \nu} \ge 0 ~\mbox{or}~ \rho-p_{r}-2p_{t} \ge 0.
\end{eqnarray}
where the sub-index $i$ represents radial $r$ or transverse $t$, respectively, whereas $~t^\mu$ and $l^\mu$ are time-like vector and null vector. In principle the origin of the above inequalities come from a purely geometric equations, namely the Raychaudhuri equations \cite{ray}. 
The Einstein fields equations relate the geometry of the space-time $R_{\mu\nu}$ with the matter content $T_{\mu\nu}$. Also, the Raychaudhuri equations can be combined to obtain the above inequalities. 
In Fig. \ref{fig6} it is clear that the SEC (Upper left panel) and DEC (Lower panels) are satisfied for all points within the compact object. Moreover, as inequalities (\ref{eq60})-(\ref{eq63}) shown the WEC is implied by DEC and NEC by SEC. Besides, as $\rho$, $p_{r}$ and $p_{t}$ are strictly positive functions in the range $0\leq r\leq R$ which ensure the fulfilment of the NEC and WEC, so we can conclude that the energy-momentum tensor threading the matter distribution of the fluid sphere is well behaved and represents an admissible physical fluid. Furthermore, the TEC condition is also satisfied as depicts the upper right panel in Fig. \ref{fig6}.

\section{Astrophysical observables}\label{section6}

The fact to include an extra piece $\theta_{\mu\nu}$ into the matter sector via a dimensionless parameter $\alpha$, alters many important properties of the compact structures. Indeed the linear mapping expressed by Eq. (\ref{deformationlambda}), introduced in formalism to separate the seed source from the new piece, induce in a natural way a modification on the gravitational mass definition. As usual, by integrating the temporal component of the Einstein field equations (\ref{effectivedensity}) one obtains
\begin{equation}\label{eq69}
m(r)=4\pi \, \int^{r}_{0} \rho(x)\, x^{2}\, dx = \frac{r}{2}\left[1-e^{-\lambda(r)}\right].
\end{equation}
However, taking into account (\ref{effecrho}) and (\ref{deformationlambda}) Eq. (\ref{eq69}) becomes
\begin{equation}\label{eq70}
m(r)=4\pi \, \int^{r}_{0} \left[\rho(x)+\alpha\theta^{t}_{t}(x)\right]\, x^{2}\, dx = \frac{r}{2}\left[1-\mu(r)-\alpha\,f(r)\right].    
\end{equation}
So, one can re--write the mass function (\ref{eq70}) as 
\begin{equation}\label{eq71}
m(r)=\underbrace{\frac{r}{2}\left[1-\mu(r)\right]}_{m_{GR}(r)}-\alpha\frac{r}{2}f(r),   
\end{equation}
that is, the gravitational mass can be seen as the usual gravitational GR mass plus a contribution coming from the MGD process. It is clear from (\ref{eq71}) that more massive compact objects can be obtained when MGD is present iff $f(r)>0$ and $\alpha<0$ (and vice versa), for all $r\in[0,R]$. In this case as was pointed out before $\alpha$ must be negative in order to ensure a positive anisotropy factor, what is more the expression for the decoupler function $f(r)$ given by (\ref{eq31}) shows that $f(r)$ is a strictly positive defined quantity. Additionally, the compactness factor $u$ also changes 
\begin{equation}\label{eq72}
2u=2u_{GR}-\alpha f(r).     
\end{equation}
Again $\alpha=0$ regains the GR case, what is more as the seed solution is isotropic the mass--radius ration in bounded by the well known Buchdhal limit \cite{buch}. In a more general context, the inclusion of new fields \cite{chakra} and anisotropies \cite{harko1} modified the mentioned limit. In this sense it is possible the so--called extra packing of mass \cite{arias} in the framework of gravitational decoupling by MGD. As tables \ref{table1} and \ref{table4} display the total gravitational mass $M$ and mass--radius ratio $u$ are greater than its GR counterpart, what is more as $\alpha$ increases in magnitude the mentioned macro physical parameters also increase in magnitude. This is so, because these observables have a linear dependence in $\alpha$. The surface gravitational red--shift $z_{s}$ is closely related with these observables. In fact $z_{s}$ is defined as follows
\begin{equation}\label{eq73}
z_{s}=\frac{1}{\sqrt{1-2u}}-1=\frac{1}{\sqrt{1-2u_{GR}+\alpha f(r)}}.  
\end{equation}
Tables \ref{table1} and \ref{table4} depict the corresponding values of this important astrophysical observable in the GR scenario and MGD one. In this direction, Ivanov \cite{iva} pointed out that any spherically symmetric anisotropic fluid sphere whose transverse pressure $p_{t}$ satisfies SEC, the surface gravitational red--shift has an upper limit given by $z_{s}=3.842$, whereas if $p_{t}$ responds to DEC the upper limit is $z_{s}=5.211$. In this opportunity, the tangential pressure satisfies both SEC and DEC, then $z_{s}$ is constrained by $5.211$. As can be seen the resulting values are within the mentioned range. In Fig. \ref{fig8} are shown the trend of $M$, $u$ and $z_{s}$ versus the dimensionless radial coordinate $r/R$.  
On the other hand, to see the stiffness of the equation of state (\ref{eos}) relating the components of $\theta$--sector, we have plotted the  M--R curve for different values placed on the tables \ref{table1}, \ref{table2}, \ref{table4} and \ref{table5} corresponding to the strange star candidates Her X--1, SMC X--1 and LMC X--4 (see Fig. \ref{fig8}). In table \ref{table8} are exhibited the corresponding numerical data extracted from Fig. \ref{fig8}. It is observed that for the constant parameters obtained for the mentioned compact stars, the maximum mass overcome the established limit for neutron and quark stars \i.e, $M=2M_{\odot}$. Nevertheless, as $\alpha$ increases in magnitude the maximum mass is less. On the other hand, the compactness factor remains bounded by the Buchdahl limit $4/9$. This fact is very important, since it says that the system will be not undergone a gravitational collapse. Respect to radius $R$ the reported values obtained from Fig. \ref{fig8} are within the branch of neutron stars \cite{abu,rawls}. In considering the surface gravitational red--shift, its values are bounded by $5.211$. Moreover, these values are within the scope of the reported numerical data in \cite{sunil1,sunil2,sunil3} for the same model.

\begin{figure}[H]
\includegraphics[width=0.32\textwidth]{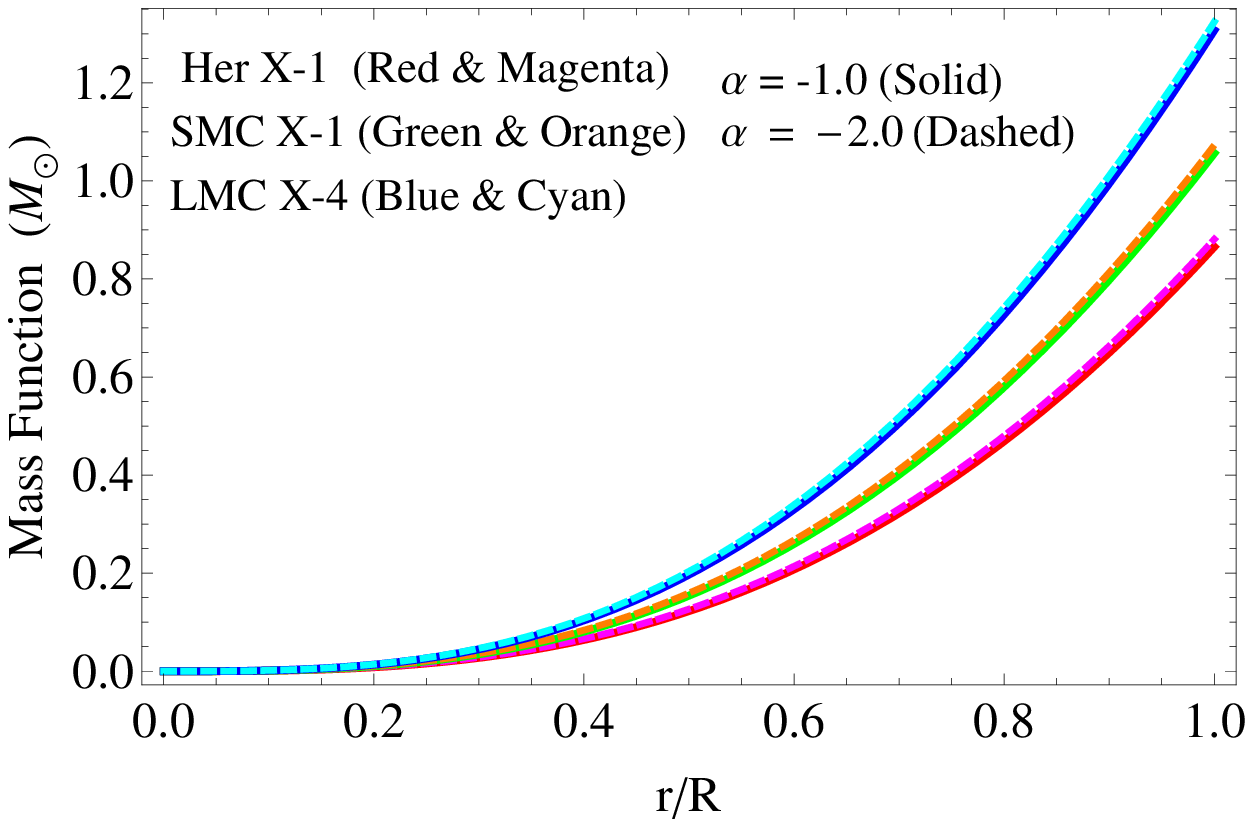} \
\includegraphics[width=0.32\textwidth]{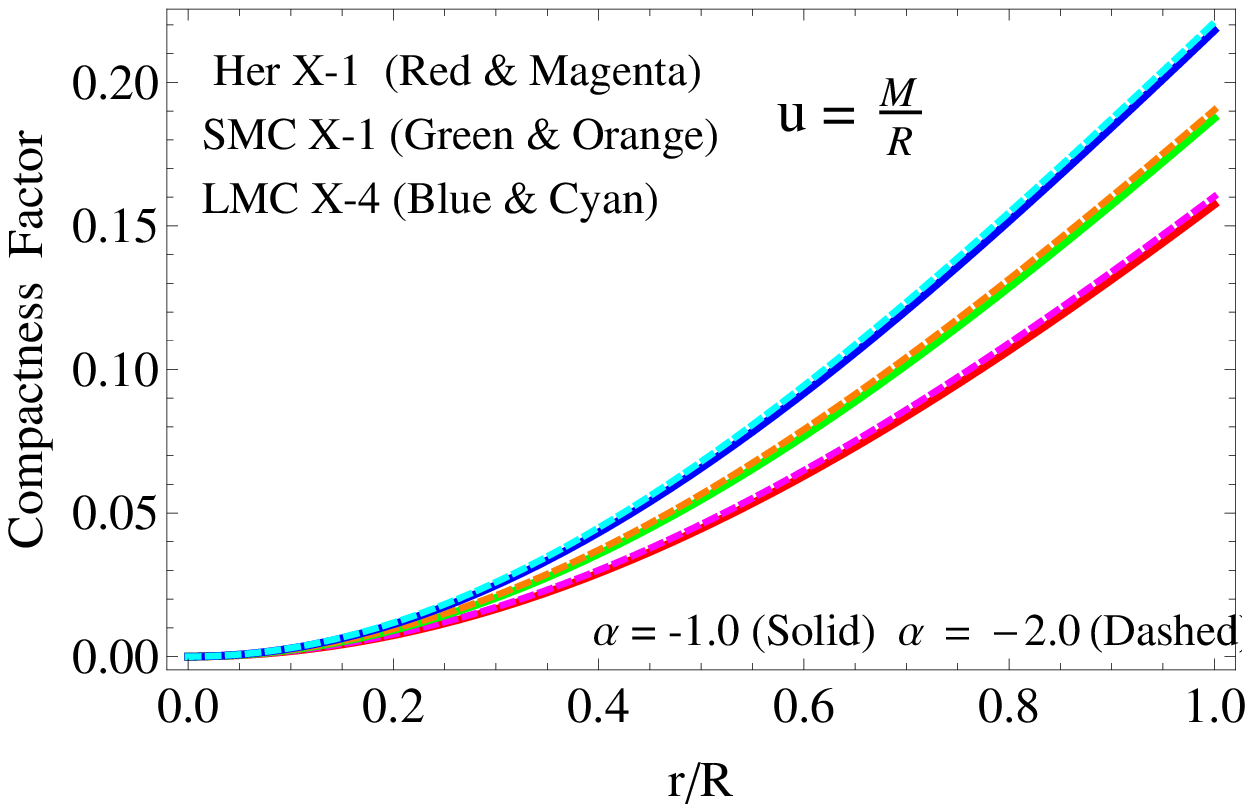}\ 
\includegraphics[width=0.32\textwidth]{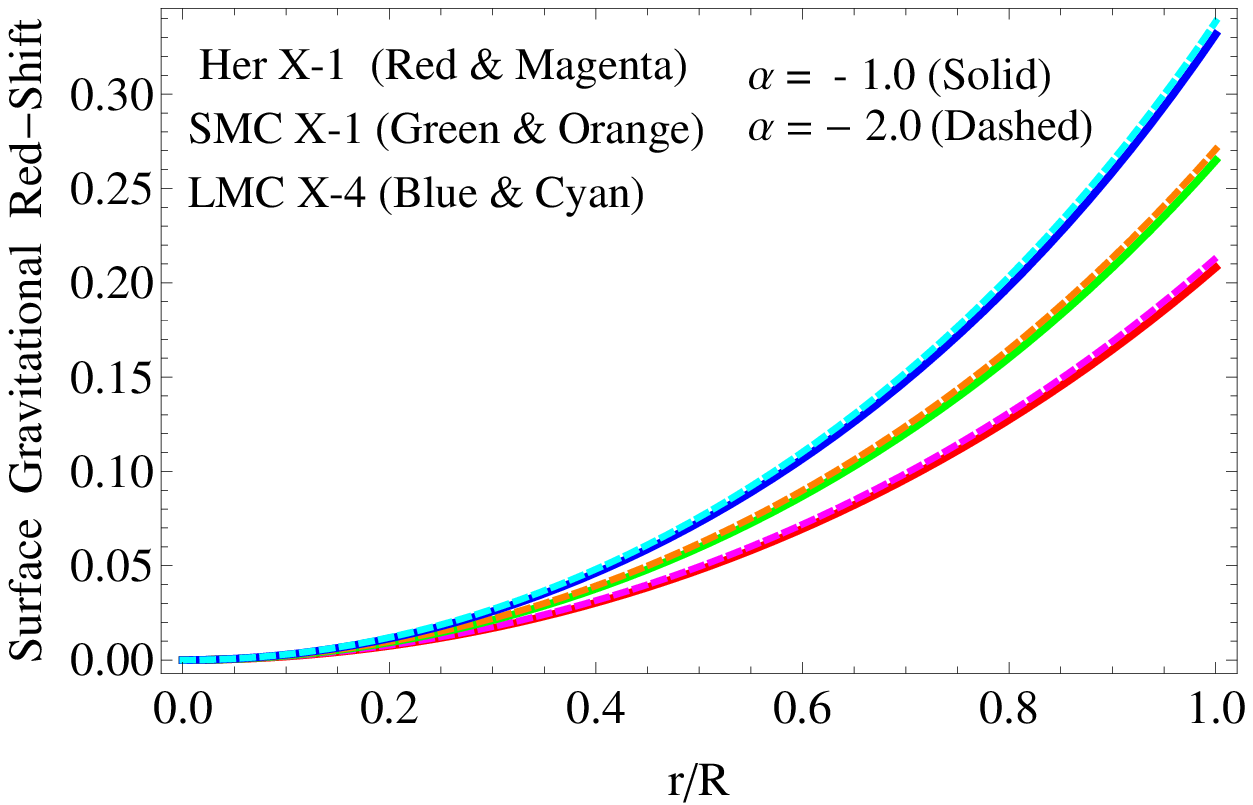}
\caption{
{\bf Left panel}: The mass function versus $r/R$. {\bf Middle panel}: The mass--radius ratio versus the dimensionless radius $r/R$.  {\bf Right panel}: The trend of the surface gravitational red--shift versus $r/R$.  }
\label{fig7}
\end{figure}

\begin{figure}[H]
\centering
\includegraphics[width=0.38\textwidth]{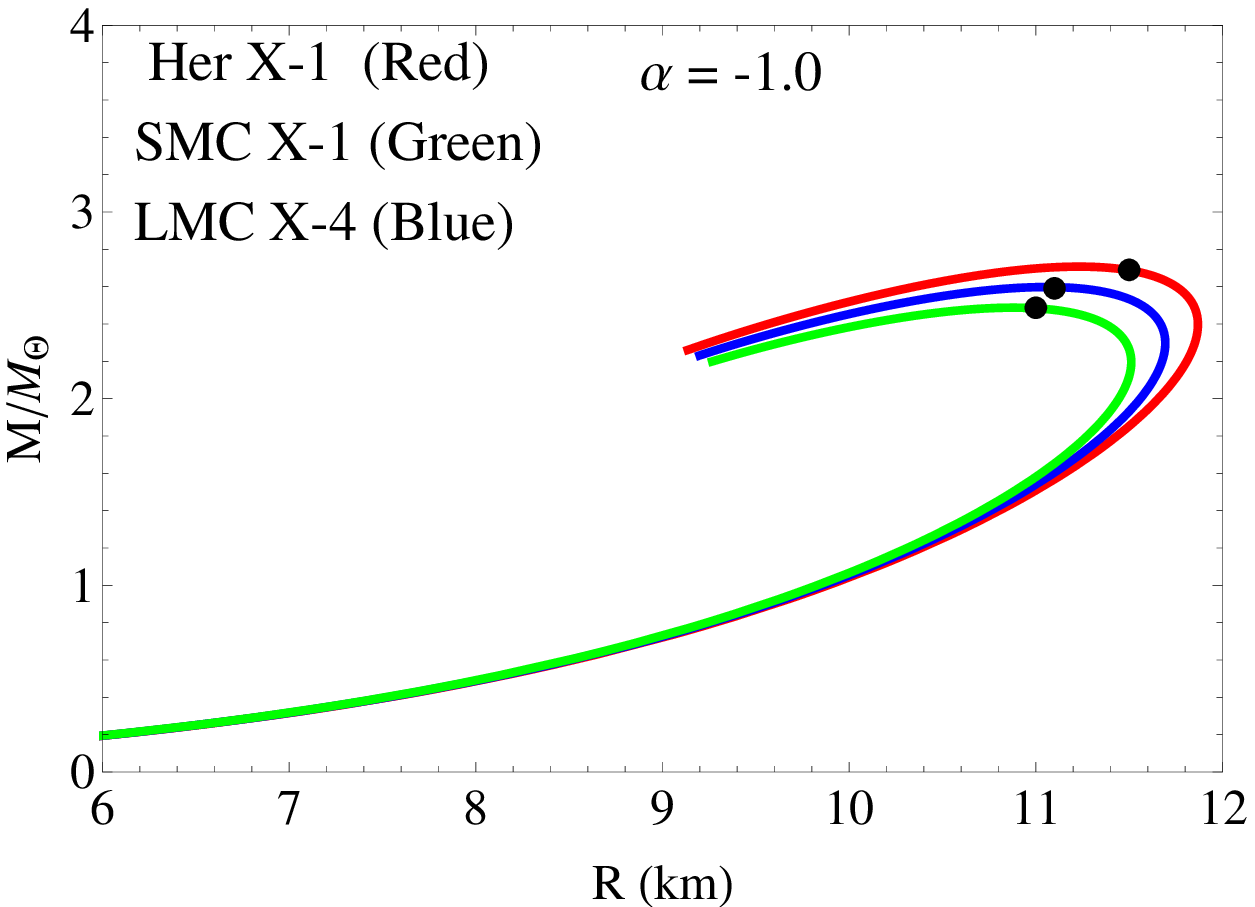} \
\includegraphics[width=0.38\textwidth]{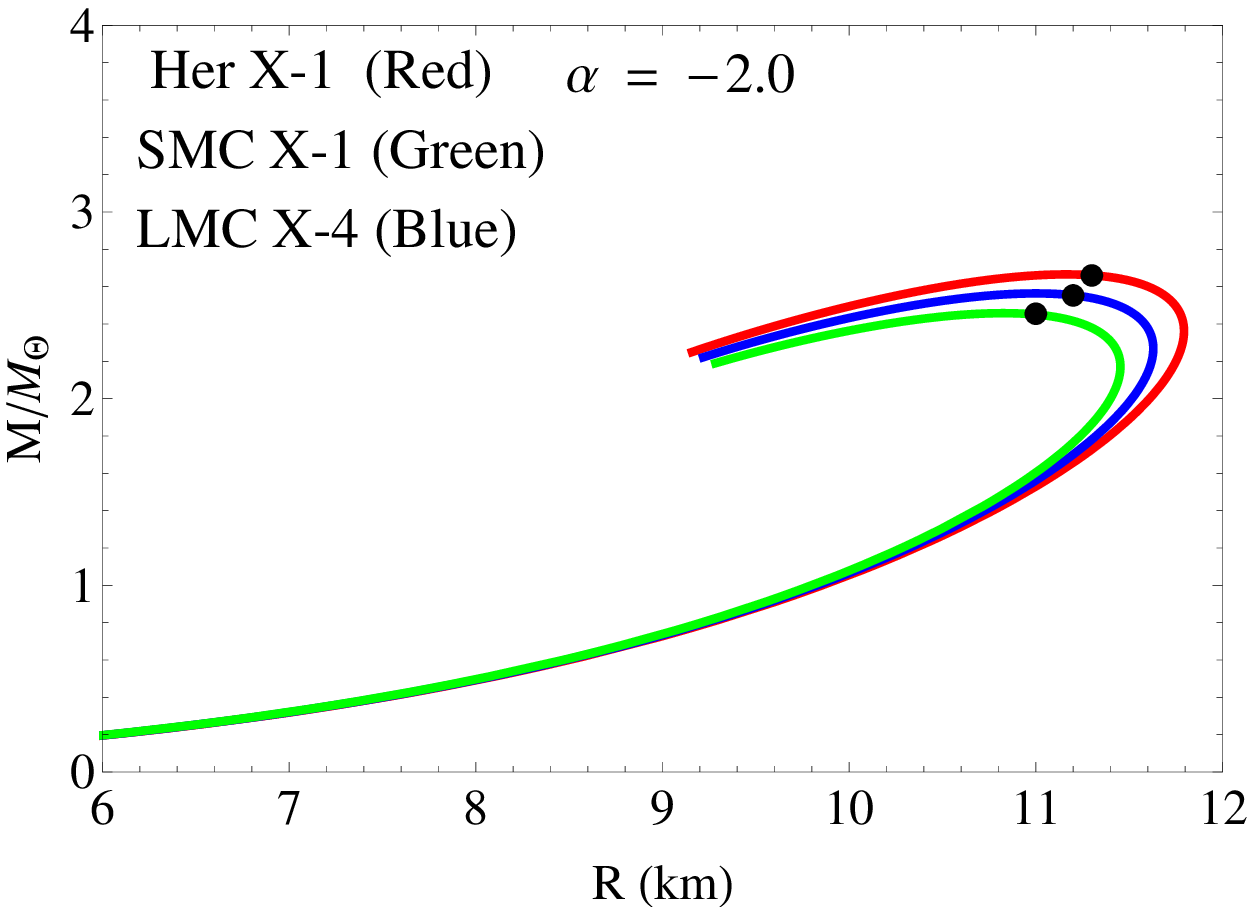}
\caption{Variation of the total mass $M$ against the radius $R$, for different values of the constant parameters mentioned in tables \ref{table1}, \ref{table2}, \ref{table4} and \ref{table5}.}
\label{fig8}
\end{figure}

\begin{table}[H]
\caption{The numerical values for the maximum mass, radius, compactness factor and surface gravitational red--shift obtained from Fig. \ref{fig8} for $a=-1$, $b=4$, different values of $\alpha$ and values mentioned in tables \ref{table2} and \ref{table5}.}
\label{table8}
\begin{tabular*}{\textwidth}{@{\extracolsep{\fill}}lrrrrrrrl@{}}
\hline
 \multicolumn{1}{c} {$\alpha$}& 
\multicolumn{1}{c} {$M_{\text{max}}/M_{\odot}$}& 
\multicolumn{1}{c} {$R\ [\text{km}]$}& 
 \multicolumn
 {1}{c} {$u=M_{\text{max}}/R$}&
 \multicolumn{1}{c}{$z_{s}$}&
  \\
\hline
 $-1$ &2.690&11.5& 0.344601&0.793747  \\
\hline
 $-1$ &2.591&11.1&0.343879&0.789593   \\
\hline
 $-1$ &2.487&11.0& 0.333077&0.730720   \\
\hline
 $-2$ &2.660&11.3& 0.346789&0.806511  \\
\hline
 $-2$ &2.553&11.2& 0.335810& 0.745066  \\
\hline
 $-2$ &2.455&11.0& 0.328791&0.708922   \\
\hline
\end{tabular*}
\end{table}

\section{Concluding remarks}\label{section7}

In the present manuscript we have investigated in detail the Durgapal's four model in light of the minimal geometric deformation approach by gravitational decoupling. In order to do it, we take advantage of a general equation of state for the anisotropic sector and, for concrete values of the parameters $a$ and $b$, we compute the corresponding deformation function as well as the components of the $\theta_{\mu \nu}$ tensor, obtaining finally the corresponding effective fluid parameters. What is more, considering that the density and the radial pressure should be monotone decreasing function from the center towards the surface of the star, we have bounded the parameter $B$ of the modified solution. This parameter plays and important role in the original solution because preservation of causality condition within the stellar interior depends on the magnitude and sign of this parameter. In this case the lower and upper bound of $B$ are altered by the presence of $\alpha$. This means that the choice of $\alpha$ is constrained to satisfy causality condition. This fact is preserved if and only if $B$ takes negative values. 
The behaviour of the thermodynamic observables was studied for different values of the parameters involved in the model. As can be seen in Fig. \ref{fig2} the trend of these variables is well behaved inside the star. Besides, to ensure a well posed stellar interior the parameter $\alpha$ must be negative. In tables \ref{table3} and \ref{table6} are shown the central density, surface density and central pressure. As can be seen the first two increase with increasing mass and $\alpha$ (in magnitude), whereas the central pressure decreases in magnitude. On the other hand, table \ref{table7} displays the same parameters for the GR case \i.e, $\alpha=0$. 
The stability of the system was checked by using different approaches: i) subliminal sound speeds of the pressure waves, ii) relativistic adiabatic index, iii) convection and iv) the Harrison--Zeldovich--Novikov criteria. In considering i) and iii) the system presents instabilities for massive objects, whereas when ii) and iv) are used the system is stable for all cases under consideration (for further details see Figs. \ref{fig3} and \ref{fig4}). The stability analysis is very important, since it determines if the hydrostatic equilibrium under different gradients is stable or not. 
In this case, the present model is under hydrostatic balance produced by the hydrostatic $F_{h}$, gravitational $F_{g}$ and anisotropic $F_{a}$ gradients. The Fig. \ref{fig5} shows the trend of these gradients along the compact structure. The positive nature of the anisotropic force $F_{a}$ induced by MGD, helps to the hydrostatic $F_{h}$ one to counteract the gravitational attraction in order to avoid the collapse of the compact configuration onto a point singularity. It is clear from panels displayed in Fig. \ref{fig5} that the anisotropic gradient increases in magnitude when $\alpha$  increases in magnitude. On the other hand, despite that the model satisfies the above criteria. It is also important to check the behavior of the full energy--momentum tensor inside the compact object. In this regard, we have studied classical energy conditions, finding that the energy--momentum tensor representing the matter distribution for this toy model is completely regular and positive defined everywhere. As Fig. \ref{fig6} illustrates, the SEC, DEC and TEC are satisfied. \\
On the other hand, the effects induced by gravitational decoupling by minimal geometric deformation scheme were explored. In this concern, it is evident that the map (\ref{deformationlambda}) alters the usual gravitational mass definition as shows Eq. (\ref{eq71}). Disturbances on the gravitational mass definition yield to modifications on the mass--radius ratio and surface gravitational red--shift. Under this particular model, to produce more massive and compact objects one needs: $f(r)>0$ and $\alpha<0$. As can be seen in tables \ref{table1} and \ref{table4}, the mentioned parameters within the arena of MGD overcome their similes in the GR background. The affections on these important astrophysical parameters is quite relevant, since by comparison the numerical values obtained in the GR case under the isotropic condition, can be distinguished from its anisotropic counterpart when local anisotropies induced by MGD grasp are present. In Fig. \ref{fig7} the gravitational mass function, compactness factor and surface gravitational red--shift are shown. Furthermore, to see the stiffness of the equation of state (\ref{eos}) employed to determine the decoupler function $f(r)$, we have built the M--R plot \ref{fig8}, for the numerical data obtained for: Her X--1, SMC X--1 and LMC X--4 reported on tables \ref{table1}, \ref{table2}, \ref{table4} and \ref{table5}. As it is observed from Fig \ref{fig8} for both cases $\alpha=-1.0$ and $\alpha=-2.0$ the maximum mass and radius are $\{2.69M_{\odot},11.5\ [km]; 2.66M_{\odot},11.3 \ [km]\}$, respectively (see table \ref{table8} for more details). These results shown that the maximum mass is above to limit $M=2M_{\odot}$ for neutron and quark stars. However, the compactness factor and red--shift are within the scope of the usual numerical data. In this regard, it is worth mentioning that the maximum compactness factor remains bounded by the well known Buchdahl limit. As final remark, one can conclude that the present toy model is able to represent and describe, at least from the theoretical point of view relativistic anisotropic compact structures.

\section*{Acknowledgments}
F. Tello--Ortiz thanks the financial support by the CONICYT PFCHA/DOCTORADO-NACIONAL/2019-21190856 and projects ANT--1756 and SEM 18--02 at the Universidad de Antofagasta, Chile. F. Tello-Ortiz thanks to the PhD program Doctorado en Física mención en Física Matemática de la Universidad de Antofagasta, Chile. \'A. R. acknowldeges DI-VRIEA for financial support through Proyecto Postdoctorado 2019 VRIEA-PUCV. P. B. is thankful to IUCAA, Pune, Government of India for providing visiting associateship. Y. Gomez--Leyton thanks the financial support by the CONICYT PFCHA/DOCTORADO-NACIONAL/2020$- $21202056. 


\appendix 

\section{Equivalence between the Eqs. (\ref{eq50}) and (\ref{eq544})} \label{A}

In this appendix we will demonstrate that Eqs. (\ref{eq50}) and (\ref{eq544}) are completely equivalents in order to determine the total mass $M$ contained by the fluid sphere. So, the starting point is the field equation (\ref{effectivedensity}) 
\begin{equation}\label{A1}
\kappa {\rho}=\frac{1}{r^2}-e^{-\lambda}\left(\frac{1}{r^2}-\frac{\lambda^{\prime}}{r}\right),   
\end{equation}
where as Eq. (\ref{effecrho}) express $\rho$ is given by
\begin{equation}\label{A2}
\rho=\tilde{\rho}+\alpha\theta^{t}_{t},  
\end{equation}
and as MGD dictates (see Eq. (\ref{deformationlambda})), $e^{-\lambda}$ reads
\begin{equation}\label{A3}
e^{-\lambda(r)}\mapsto \mu(r)+\alpha f(r).   
\end{equation}
Next, by replacing (\ref{A2}) and (\ref{A3}) into (\ref{A1}) one arrives to
\begin{equation}\label{A4}
8\pi\left(\tilde{\rho}+\alpha\theta^{t}_{t}\right)=\frac{1}{r^{2}}\left[1-\mu-r\mu^{\prime}-\alpha f-\alpha rf^{\prime}\right],    
\end{equation}
where the left hand side can be re-organized as follows
\begin{equation}\label{A5}
8\pi\left(\tilde{\rho}+\alpha\theta^{t}_{t}\right)=\frac{1}{r^{2}}\bigg[r-\left(\mu+\alpha f\right)r\bigg]^{\prime}.  
\end{equation}

Now, integrating both sides of (\ref{A5}) over the volume $dV\equiv4\pi r^{2}dr$ with $r\in[0,R]$ and applying the first fundamental theorem of calculus, one obtains
\begin{equation}\label{A6}
4\pi\int^{R}_{0}\left(\tilde{\rho}+\alpha\theta^{t}_{t}\right)r^{2}dr=\frac{R}{2}\bigg[1-\mu(R)-\alpha f(R)\bigg].   
\end{equation}

On the other hand from Eq. (\ref{eq50}) 
\begin{equation}\label{A7}
\begin{split}
\frac{7-10CR^{2}-C^{2}R^{4}}{7\left(1+CR^{2}\right)^{2}}
+ \frac{BCR^{2}}{\left(1+CR^{2}\right)^{2}\left(1+5CR^{2}\right)^{\frac{2}{5}}}
+ \frac{\alpha CR^{2}}{4\left(9CR^{2}+1\right)^{2}\left(CR^{2}+1\right)^{2}}
= 1-2\frac{{M}}{R},   
\end{split}    
\end{equation}
 after some algebra, one gets the mass $M$ as follows
\begin{equation}\label{A8}
\begin{split}
M=\frac{R}{2}\bigg[1-\frac{7-10CR^{2}-C^{2}R^{4}}{7\left(1+CR^{2}\right)^{2}}
- \frac{BCR^{2}}{\left(1+CR^{2}\right)^{2}\left(1+5CR^{2}\right)^{\frac{2}{5}}}-\frac{\alpha CR^{2}}{4\left(9CR^{2}+1\right)^{2}\left(CR^{2}+1\right)^{2}}\bigg],
\end{split}    
\end{equation}
where
\begin{eqnarray}\label{A9}
\mu(R)&\equiv& \frac{7-10CR^{2}-C^{2}R^{4}}{7\left(1+CR^{2}\right)^{2}}
 - \frac{BCR^{2}}{\left(1+CR^{2}\right)^{2}\left(1+5CR^{2}\right)^{\frac{2}{5}}}, \\ \label{A10}
 f(R)&\equiv& \frac{ CR^{2}}{4\left(9CR^{2}+1\right)^{2}\left(CR^{2}+1\right)^{2}}.
\end{eqnarray}
Therefore, as can be seen Eqs. (\ref{eq50}) and (\ref{eq544}) lead to the same expression to compute the total mass $M$ inside the compact object. As final remark, it should be noted that $\alpha=0$ leads to the general relativity gravitational mass definition, although as usual one can recast the general relativity expression by rewriting $\tilde{\rho}+\alpha\theta^{t}_{t}=\rho$ and $\mu(r) + \alpha f(r)=e^{-\lambda}$ \i.e, by using the effective quantities. Nevertheless, it must be considered that the density $\rho$ and metric potential $e^{-\lambda}$ of the solution in that appearance hide the contribution from gravitational decoupling and MGD. 

\section{The generating function} \label{B}

As was pointed out by Herrera et al. \cite{gene}, all the spherically symmetric and static models whose matter distribution is described by an imperfect fluid (see \cite{cha} for the charged case), can be acquired from 
two generating functions. These
two primitive generating functions $\zeta(r)$ and $\Pi(r)$ are given by
\begin{equation}\label{B1}
e^{\nu(r)}=\text{Exp}\left[\int\left(\zeta(r)-\frac{2}{r}\right)dr\right] \Rightarrow \zeta(r)=\frac{\nu^{\prime}(r)}{2}+\frac{1}{r},   
\end{equation}
\begin{equation}\label{B2}
\Pi(r)=\left(p_{r}-p_{t}\right)=-\Delta(r).    
\end{equation}
Then, by using Eqs. (\ref{eq2}), (\ref{eq39}) and (\ref{B1})--(\ref{B2}) one arrives at the following generators
\begin{equation}\label{B3}
\zeta(r)=4AC\left(1+Cr^{2}\right)^{3}r+\frac{1}{r},    
\end{equation}
\begin{equation}\label{B4}
\Pi(r)= \frac{\alpha\left(9Cr^{2}+2\right)C^{2}r^{2}}{4\pi\left(9Cr^{2}+1\right)^{3}\left(Cr^{2}+1\right)^{3}}.   
\end{equation}
At this point it should be noted that, the second generator (\ref{B2}) is naturally induced by gravitational decoupling by means of MGD. So, if $\alpha=0$, the solution turn on isotropic (one recovers the seed solution in the GR arena), then (\ref{B4}) is null. In that case the generator is given only by (\ref{B1}), coinciding with the results reported by Lake in \cite{lake11}.

\end{document}